\documentclass[12pt]{article}
\usepackage[margin=2 cm]{geometry}
\usepackage{comment}
\usepackage{graphicx}
\usepackage{tikz}
\usetikzlibrary{arrows}
\usepackage{mathtools}
\usepackage[font={small,it}]{caption}
\usepackage{mwe}
\usepackage[nottoc]{tocbibind}
\usepackage{amsmath,amssymb,extarrows,mathtools,graphicx,subfigure,setspace}
\usepackage{cite}
\usepackage{braket}
\usepackage{epsfig}
\usepackage[section]{placeins}
\usepackage{slashed}
\usepackage{color}
\usepackage{xcolor}
\usepackage{caption}
\usepackage{amsmath}
\usepackage[implicit]{hyperref}
\makeatother

\newcommand{\be}{\begin{equation}}
\newcommand{\bea}{\begin{eqnarray}}
\newcommand{\eea}{\end{eqnarray}}
\newcommand{\ba}{\begin{array}}
\newcommand{\ea}{\end{array}}
\newcommand{\ee}{\end{equation}}
\newcommand{\bes}{\begin{equation*}}
\newcommand{\beas}{\begin{eqnarray*}}
\newcommand{\eeas}{\end{eqnarray*}}
\newcommand{\bas}{\begin{array*}}
\newcommand{\eas}{\end{array*}}
\newcommand{\ees}{\end{equation*}}

\numberwithin{equation}{section}

\begin{document}

\onehalfspacing
\vfill
\begin{titlepage}
\vspace{10mm}

\begin{center}

\vspace*{10mm}
\vspace*{1mm}
{\Large  \textbf{On complexity growth in massive gravity theories, the effects of chirality and more }} 
 \vspace*{1cm}
 
{$\text{Mahdis Ghodrati}^{a}$}

\vspace*{8mm}
{ \textsl{
$^a $School of Particles and Accelerators,
Institute for Research in Fundamental Sciences (IPM) \\
P.O. Box 19395-5531, Tehran, Iran}} 
 \vspace*{1cm}

\textsl{E-mails: {\href{mailto:mahdisghodrati@ipm.ir}{mahdisghodrati@ipm.ir}}}
 \vspace*{2mm}

\vspace*{1.7cm}

\end{center}

\begin{abstract}

To study the effect of parity-violation on the rate of complexity growth, by using ``Complexity=Action" conjecture, we find the complexity growth rates in different solutions of the chiral theory of Topologically Massive Gravity (TMG) and parity-preserving theory of New Massive Gravity (NMG). Using the results, one can see that decreasing the parameter $\mu$, which increases the effect of Chern-Simons term and increases chirality, would increase the rate of growth of complexity. Also one can observe a stronger correlation between complexity growth and temperature rather than complexity growth and entropy. At the end we comment on the possible meaning of the deforming term of chiral Liouville action for the rate of complexity growth of warped CFTs in the Tensor Network Renormalization picture.
 \end{abstract}

\end{titlepage}

\tableofcontents


\section{Introduction}
Based on AdS/CFT duality and holography, one should be able to calculate different parameters of a boundary CFT by using the dual bulk theory. One such quantity is the complexity of a quantum state where in quantum information context is defined by using the minimum number of simple gates which are needed to build a quantum circuit that constructs them from a certain reference state \cite{Watrous:2008}.  There are also some recent progresses in \cite{Chapman:2017rqy, Jefferson:2017sdb} to define complexity more rigorously in quantum field theory and in a continuous way, where interestingly their results in different setups match with results from holography. 

The holographic proposal, by Susskind  \cite{Susskind:2014rva, Brown:2015bva }, states that for computing the quantum computational complexity of a holographic state one can calculate the on-shell action on the ``Wheeler-De Witt'' patch. Therefore,
\begin{gather}\label{eq:defA}
\mathcal{C} \left( \Sigma \right) =\frac{ I_ {WDW}}{\pi \hbar},
\end{gather}
where $\Sigma$ is the time slice which is the intersection of asymptotic boundary and the Cauchy surface in the bulk. This proposal is named Complexity=Action (CA) conjecture.

There is also Complexity=Volume (CV) conjecture \cite{Susskind:2014rva} which states that to compute the complexity of the boundary state, one can evaluate the volume of a codimension-one bulk hypersurface intersecting with the asymptotic boundary on the desired time slice. So
\begin{gather}\label{eq:defv}
\mathcal{C}_\mathcal{V} \left( \Sigma \right) = {\text{max}}_{\Sigma=\partial \mathcal{B}} \left[ \frac{\mathcal{V}(\mathcal{B})}{G_N \ell} \right],
\end{gather}
where $\mathcal{B}$ is in the bulk and $\ell$ is a specific time-scale such as the radius of AdS space.

The complexity grows linearly even after the boundary reaches the thermal equilibrium and therefore it could be a useful thermodynamical and quantum information measure. In the dual picture, this growth of complexity corresponds to the expansion of the length of Einstein-Rosen bridge (ERB) or the volume of the wormhole entangling two thermofield-double CFTs on boundary. Studying properties of complexity could also help to understand the inside of black holes.

 In \cite{Lloyd:2000}  an upper bound for the rate of growth of quantum complexity has been found and later in \cite{Brown:2015lvg} it was written in the holographic context as
 \begin{gather}\label{eq:bound}
\frac{d \mathcal{C}}{dt} \le \frac{2M}{\pi \hbar},
\end{gather}
where $M$ is the mass of the black hole in the bulk, and this inequality saturates for the uncharged black holes.

\begin{figure}
 \centering
  \includegraphics[width=5cm] {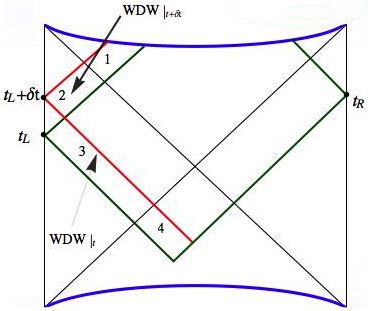}
  \includegraphics[width=4cm] {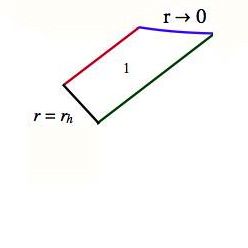}
  \caption{General Penrose diagram and the corresponding WDW patch for calculating complexity growth. At the late time only region 1 contributes to the complexity growth.}
 \label{fig:gp}
\end{figure}

In \cite{Alishahiha:2017hwg}, using the CA conjecture, by calculating the on-shell actions on two nearby WDW patches, shown in Fig. \ref{fig:gp}, the rates of complexity growth for gravity theories with higher derivative terms, such as $F( R  )$ and New massive Gravity (NMG), for specific black hole solutions and shockwave have been calculated and the above bound for complexity growth rate has been verified.

Those theories, however, are parity preserving having both left and right moving modes in the dual boundary CFTs. In this work, we are mainly interested in studying the effect of chirality on the rate of growth of complexity. Notably, the effects of chirality on the entanglement entropy in parity-violating quantum filed theories have been studied in \cite{Hughes:2015ora, Iqbal:2015vka, Castro:2015csg}. There, an \textit{entanglement inflow} between the bulk and the domain-wall has been perceived which actually comes from the imbalance in the flux of modes flowing through the boundary.  So, it would be very interesting to check if such effects can also be detected by calculating the holographic complexity of the bulk in parity-violating gravitation theories and specifically to study the effects of edge states.

In this work, first we study the effect of Chern-Simons term on the rate of growth of complexity. Again using the CA conjecture, we calculate the rate of complexity growth in several solutions of \textit{Topologically Massive Gravity (TMG)} which is the Einstein-Hilbert action plus the chiral breaking Chern-Simons term.

As mentioned in \cite{Alishahiha:2017hwg}, the main challenge is to calculate the contribution coming from the boundary term. For the Gibbons-Hawking boundary term of TMG we specifically use the boundary term first introduced in \cite{Miskovic:2009kr} where the background-independent charges of TMG have been calculated. Considering that the approach of \cite{Brown:2015bva, Brown:2015lvg} has worked for \cite{Alishahiha:2017hwg}, we go forward and use it for different black hole solutions  of TMG namely, BTZ, warped $\text{AdS}_3$, Null Warped $\text{AdS}_3$, supersymmetric and ACL black holes.  We will also present the result for the shockwave solution of TMG in our following paper.  

For the sake of comparing our results with the parity-preserving case, we also calculate complexity growth in warped $\text{AdS}_3$, new hairy and log black hole solutions of NMG and comment on the effect of warping factor, hair parameter and log term on the growth rate of complexity. We also compare  complexity growth rate with different thermodynamical quantities of these black holes and and observe a curious correlation between temperature and complexity growth which might be useful in understanding thermodynamical-like laws for complexity. Finally, we conclude with a discussion where we comment on many recent progresses in defining quantum complexity in CFTs which one could also apply for the warped CFT case as well. Specifically, we compare the usual Liouville and ``chiral Liouville" actions to try to interpret the meaning of the warped CFT deformed term in the MERA language.

\section{Complexity growth in a chiral theory}

The chiral theory of topologically massive gravity, also known as Chern-Simons gravity is a rich, ghost-free theory of gravity in $2+1$ dimensions. The field equations of this theory include the Cotton tensor which is the analogue of the Weyl tensor in three dimensions and it can add a degree of freedom to the theory to make it dynamical which also makes the graviton massive. The effects of all these could change the rate of complexity growth.

In first order formalism, the action of TMG with a negative cosmological constant $\Lambda= - 1/\ell^2$ can be written as \cite{Miskovic:2009kr}
\begin{gather}\label{eq:TMGaction}
I= - \frac{1}{ 16 \pi G} \int_M \epsilon_{ABC} \left ( R^{AB} + \frac{1}{3 \ell^2} e^A e^B \right) e^C+ \frac{1}{32 \pi G \mu} \int_M \left( L_{CS} \left (\omega \right) + 2 \lambda_A T^A \right)+\int _{\partial M} B.
\end{gather} 

In the above action, $M$ is a three-dimensional manifold where $x^\mu$ are the local coordinates, $G$ is the gravitation constant, $\mu$ is a constant parameter with the dimension of mass and $L_{CS}$ is the gravitational Chern-Simons 3-form which its relation is
\begin{gather}
L_{CS} (\omega) = \omega ^{AB} d \omega_{BA}+\frac{2}{3} {\omega^ A}_B {\omega^B}_C {\omega^C}_A.
\end{gather}

By defining the dreibein $e^A= e^A_\mu dx^\mu $ and the spin connections $ \omega^{AB} =\omega_\mu ^{AB} dx^\mu $ one can write the curvature 2-form as
\begin{gather}
 R^{AB} =\frac{1}{2} R_{\mu\nu}^{AB} dx^\mu dx^\nu = d \omega^{AB}+ {\omega^A}_C \omega^{CB},  
\end{gather}
and then the torsion 2-form as $T^A=\frac{1}{2} T^A_{\mu \nu} dx^\mu dx^\nu = De^A$, where the covariant derivative acts on the vectors as $DV^A= dV^A+{\omega^A}_B V^B$. As one would want a torsionless theory, then $T_A=0$ and then one can find the Lagrange multipliers in terms of Schouten tensor of the manifold
\begin{gather}
S_{\mu \nu}= (Ric)_{\mu\nu}-\frac{1}{4} \mathcal{G}_{\mu \nu} R,
\end{gather} 
as
\begin{gather}
\lambda_mu^A=-2 e^{A\nu} S_{\mu \nu},
\end{gather}
where $\mathcal{G}=\eta_{AB} e_\mu^A e_\nu^B$.

For the first time, for the TMG case, in \cite{Miskovic:2009kr}, the boundary term which makes the variational principle well-defined were introduced as 
\begin{gather} \label{eq:boundary}
B= \frac{1}{32\pi G} \epsilon_{ABC} \omega^{AB} e^C.
\end{gather}
Note that specially for the topological theories and Chern-Simons action the contribution of the boundary term is significant as it is also the case for the modes on the boundary of topological matters. 

Now, as explained in \cite{Bouchareb:2007yx}, TMG admits two different kinds of black hole solutions, one is asymptotically AdS, BTZ solution of Einstein gravity with a negative cosmological constant, and the other is the non-asymptotically flat, non-asymptotically AdS, with a zero cosmological constant ACL black hole.
In the following sections we calculate the rate of complexity growth for these two categories of black holes and study the effect of different parameters of the theory and solutions, specifically the parameter $\mu$ on this growth rate.

\subsection{BTZ black hole}

By the method introduced in \cite{Alishahiha:2015rta, Alishahiha:2017hwg }, we evaluate the TMG action for the BTZ case.   For the BTZ metric of 
\begin{gather}
ds^2= -f (r )  ^2 dt^2 +\frac{dr^2}{f(r )  ^2} +r^2 (d\phi-\frac{4G J}{r^2} dt )^2,  \ \ \ \ \  f^2(r  ) = \frac{r^2}{\ell^2}-8GM+\frac{(8GJ)^2}{4r^2}.
\end{gather} 
the vierbeins and spin connections would be \cite{Bouchareb:2007yx}
\begin{gather}
e^0= f( r  ) dt, \ \ \ \ \ \ \ \ \ e^1= (r d\phi -\frac{4G J}{r}) dt, \ \ \ \ \ \ \ \ \ e^2=\frac{1}{f( r )} dr, \nonumber\\
{\omega^0}_1 =\frac{4 G J}{r^2  f( r )} dr, \ \ \ \ \ \  {\omega^0}_2= \left( f'( r ) f( r ) -\frac{16 G^2 J^2}{r^3} \right) dt+\frac{4GJ}{r} d\phi, \ \ \ \ \ \  {\omega^1}_2= f( r ) d\phi. 
\end{gather}

\begin{figure}
 \centering
  \includegraphics[width=8.5cm] {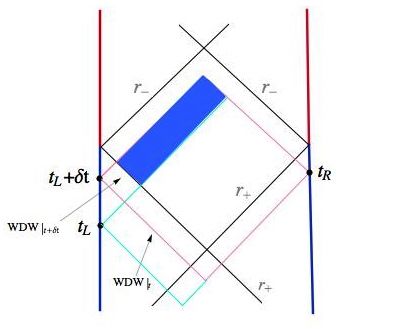}
    \caption{ Penrose diagram of BTZ black hole. At late times, only the dark blue part contributes to the complexity growth.}
  \label{fig:BTZpenrose}
\end{figure}

Note that for the BTZ case, the Cotton tensor vanishes identically and so it satisfies the TMG field equations in a trivial way. 
Now calculating the Lagrangian, the first term, $\epsilon_{ABC} R^{AB} e^C$, gives
\begin{gather}
\epsilon_{ABC} R^{AB} e^C=  2\left( 2 f'(  r ) f ( r) +r f'' ( r ) f ( r) + r f'^2 (  r  ) +\frac{4 G^2  J^2}{r^3}    \right) dt dr d\phi.
\end{gather}
For the second term we get
\begin{gather}
\frac{1}{3 \ell^2} \epsilon_{ABC} e^A e^B e^C = -\frac{2r}{\ell^2} dt dr d\phi. 
\end{gather}
Also for the BTZ metric, the Chern-Simon term would give
\begin{gather}
L_{CS}= -\frac{8 G J}{r} \left( \frac{64 G^2 J^2}{r^4} + f'' f +f'^2-\frac{f' f}{r} \right).
\end{gather}
One can also check that as the Lagrange multiplier for the locally AdS space is  $\lambda_\mu^A=\frac{1}{\ell^2} e_\mu^A$ then $\lambda_A T^A=0$ and there would be no contribution from this term as one expects from the equations of motion of TMG. Also for the boundary term $B$ one finds
\begin{gather}
B=\frac{1}{32 \pi G} \epsilon_{ABC} \omega^{AB} e^C= 2 f \left(f + r f' \right)  d\phi \wedge dt.
\end{gather}

Now we can write the parameters of the BTZ metric in terms of the outer and inner horizon radii, $r_{+} \ , r_{-}$ (the solutions of $f( r  )=0$) in the following form,
\begin{gather}
f^2(  r ) =\frac{(r^2 -r_+^2)(r^2- r_-^2) }{ r^2 \ell^2}, \ \ \ \ \ 8GM=\frac{r_+^2 +r_-^2}{\ell^2}
, \ \ \ \ \ \ \  8GJ=\frac{2 r_+ r_-}{\ell}.
\end{gather}
Also the total mass and total angular momentum of TMG could be written as \cite{Miskovic:2009kr}
\begin{gather}
\mathcal{M} =M-\frac{J}{\mu \ell^2}, \  \ \ \ \ \ \ \ \ \ \  \mathcal{J}=J-\frac{M}{\mu}.
\end{gather}

Now similar to \cite{Alishahiha:2017hwg}, to find the rate of the growth of complexity, one should calculate the difference between the on-shell actions which are evaluated over the two nearby WDW patches. At the late time the only part that contributes to the rate of complexity growth is region 1 which is shown in blue in Figure. \ref{fig:BTZpenrose}. For the BTZ case and at the late time, only the region between the two horizons contribute to this difference. So one would find
\begin{flalign}
\delta I_{\mathcal{M}} &= I_{\mathcal{M}} [ \text{WDW} \big |_{t +\delta t} ] - I_{\mathcal{M}} [\text{WDW}\big |_t ] \nonumber\\ &= -\frac{1}{16 \pi G} \int_{r_-}^{r_+} \int_{t} ^{t+\delta t} \int_{0}^{2\pi} \mathcal{L_{\text{EH}}}  \ dt \  dr \  d\phi + \frac{1}{32 \pi G \mu} \int _{r_-}^{r_+} \int_t ^{t+\delta t} \int_0^{2\pi} L_{\text{CS}}   \ dt \  dr \  d\phi  \nonumber\\&=
-\frac{\delta t}{4 G \ell^2} \int_{r_-}^{r_+} \left(r+\frac{r_+^2 r_-^2}{r^3} \right) dr  - \frac{\delta t J}{2 \mu} \int_{r_-} ^{r_+} \frac{dr}{r} \left( \frac{64 G^2 J^2}{r^4} + f'' f + f'^2 -\frac{f' f}{r} \right) \nonumber\\
&= -\frac{(r_+^2 -r_-^2)}{4G \ell^2}\delta t+ \frac{1}{4 G \ell^3 \mu} \left( \frac{r_+^4 -r_-^4}{r_+ r_-} \right)    \delta t.
\end{flalign}
The first term coming from the Einstein Hilbert term, matches with previous calculations such as in\cite{Alishahiha:2017hwg}.

Then the contribution of the generalized Gibbons-Hawking boundary term \eqref{eq:boundary} would be
\begin{flalign}
\delta I_{\partial \mathcal{M}} &= \int_t ^{t+\delta t} \int_0^{2\pi}  \frac{2}{\ell^2} \left(2r -r_+^2 -r_-^2  \right) dt \ d\phi \Big |_{ r_+}  \nonumber\\ & -  \int_t ^{t+\delta t} \int_0^{2\pi} \left(2r -r_+^2 -r_-^2  \right) dt \ d\phi \Big | _{ r_-}=\frac{(r_+^2 -r_-^2)}{4 G \ell^2} \delta t.
\end{flalign}

Based on \eqref{eq:defA}, the complexity growth would be
\begin{gather}\label{eq:cbtz}
\dot{\mathcal{C}} =\frac{d I}{dt}= \frac{1}{4 G \ell^3 \mu} \left( \frac{r_+^4 -r_-^4}{r_+ r_-} \right).
\end{gather} 
We can also write the complexity growth $\dot{\mathcal{C}}$ in terms of the conserved charges of BTZ  as
\begin{gather} \label{eq:TMGcomplex}
\dot{\mathcal{C}}=\frac{4 M}{\mu J} \sqrt{M^2 -\frac{J^2}{\ell^2} }=\frac{4}{\ell^2}\frac{\mathcal{M} \mu  \ \ell^2+ \mathcal{J} }{\mathcal{M}+\mu \mathcal{J} } \sqrt{\frac{\ell^2 \mathcal{M}^2-\mathcal{J}^2}{ \mu ^2 \ell^2 -1}}.
\end{gather}
One can notice that the higher derivative corrections which here is the Chern-Simons term, would actually slow down the rate of growth of complexity similar to the results of \cite{Alishahiha:2017hwg} for the critical gravity where the mass term decreased the rate.

 Also note that for the special case of $\mu \ell=1$ the complexity growth rate diverges indicating again that this is a special point in the region of the solution. In this critical point, the left central charge would vanish and the equation of motion degenerate to a log-gravity which its dual is the LCFT \cite{Grumiller:2008qz}. 
 
From the result of \eqref{eq:cbtz} one can see that decreasing the coupling $\mu$ which increases the effect of Chern-Simons term in the action \eqref{eq:TMGaction} and increases the parity-violation, would increase the rate of complexity growth. This actually makes sense, since breaking the symmetry between the left and right moving modes should definitely increase complexity and its growth rate. Note that for $\mu \to 0$, where the Chern-Simons term becomes completely dominant, the rate of complexity growth diverges which however might not physically be possible due to the bound of \eqref{eq:bound}. 
 
A peculiar feature of this result is that for $\mu \to \infty$ it will not give the complexity growth of pure Einstein action. This might be due to specific feature of the Chern-Simons theory, or the effect of the particular boundary term \eqref{eq:boundary} that we have chose, which is independent of the factor $\mu$, unlike NMG which depends on $m^2$ through an auxiliary field.  It worths to work further on this point and to check how actually the distinctions between the left and right moving modes and increasing chirality would increase complexity growth rate.

One might also try to interpret the results based on the difference between the central charges, 
\begin{gather}
c_L=\frac{3 \ell}{2 G} \left( 1-\frac{1}{\mu \ell} \right), \ \ \ \ \ \ \ \ \ \ c_R=\frac{3 \ell}{2 G} \left(1+\frac{1}{\mu \ell} \right ),\ \ \ \ \ \ \Delta c=\frac{3}{\mu \ell}.
\end{gather}
Note also that as in TMG case mass, i.e., $M$, could be negative, the bound of $\mathcal{\dot{C}}\le 2\mathcal{M}$ at $\mathcal{J}=0$, could be satisfied. 

To examine the behavior of complexity, we compare it with other thermodynamical quantities of the BTZ black holes in TMG which are as follows \cite{Zhang:2013mva, Myung:2008ey},
\begin{gather}\label{eq:btzthermo}
S=\frac{\pi r_+}{2G}+\frac{1}{\mu \ell} \frac{\pi r_-}{2G},  \ \ \ \ \ \ \ T_H=\frac{r_+^2 -r_-^2}{2 \pi \ell^2 r_+}, \ \ \ \ \ M=\frac{r_+^2+r_-^2}{8 H \ell^2}, \ \ \ \ \ \ J=\frac{2 r_+ r_-}{8 G \ell}.
\end{gather}
Note that for the extremal case where $T \to 0$ and $r_+ -r_- \to 0$, we have $ \dot{\mathcal{C}} \to 0$ as we expected. 

In fact, there are evidences that complexity is a quantity which shows similarities to both temperature and entropy. However, from \eqref{eq:btzthermo}, one can notice the more similarities are actually between complexity and temperature where both are always proportional to $(r_+-r_-)$. In \cite{Flory:2017ftd, Roy:2017uar} also it was shown that in certain systems by decreasing temperature complexity would decrease.  All these observations could suggest that a more direct relationship between complexity and temperature exists, rather than complexity and entropy. 

This is actually in accordance with the Lloyd's proposal in \cite{Lloyd:2000}. As he put forward, integrating $T=(\partial S/ \partial E)^{-1}$, leading to $T= CE/S$ ($C$ is just constant), suggests that the temperature would actually govern the number of operations per bit per second, $ \left( k_B \ln 2E/ \hbar S \approx k_B T/ \hbar \right)$, that a system can perform, and conceptually this is more related to the concept of complexity than entropy. By calculating the complexity of black holes, we also see this interconnection directly.

\subsection{Warped $\text{AdS}_3$ black hole}

Now we want to calculate the holographic quantum complexity for a warped $\text{AdS}_3$ black hole and study the effect of warping factor in addition to the effect of chirality. 

Warped $\text{AdS}_3$ black holes are in fact stretched or squeezed deformation of BTZ black holes. Their isometry group is $SL(2,R) \times U(1)$ and their dual boundary theory is warped CFT (WCFT) which is a semi-direct product of a Virasoro algebra and a $U(1)$ Kac-Moody algebra. 

  The metric of $\text{WAdS}_3$ black hole in the ADM form can be written as
\begin{gather}\label{eq:warped}
ds^2= \ell^2 \left( dt^2 + 2 M( r  ) dt d\theta + N ( r ) d \theta^2 + D ( r ) dr^2 \right ),
\end{gather}
 where
 \begin{flalign}
 M( r ) &= \nu r -\frac{1}{2} \sqrt{r_+ r_- (\nu^2+3)}, \nonumber\\
 N(  r ) &= \frac{r}{4}  \left(3 (\nu^2-1) r + ( \nu^2+3) ( r_+ + r_-) -4\nu \sqrt{r_+ r_- (\nu^2+3)} \right), \nonumber\\
 D( r )&= \frac{1}{(\nu^2+3)(r-r_+)(r-r_-)}. 
 \end{flalign}
Note that $ \nu= \frac{\mu \ell}{3}$ and for the case of $\nu=1$ this metric reaches to the BTZ black hole in an unusual coordinate.

The Carter-Penrose diagrams of these kinds of black holes have been presented in \cite{Ferreira:2013zta} which are similar to the asymptotically flat space times in $3+1$ dimensions. Also, in\cite{Ferreira:2013zta} it was shown that these black holes are stable against the massive scalar field perturbations.

If we choose the following vierbein  \cite{Chen:2009hg}
\begin{gather}
e^0= \frac{\ell}{2 \sqrt{D ( r )} } d\theta, \ \ \ \ \ e^1= \ell \sqrt{ D ( r )} dr, \ \ \ \ \  e^2= \ell dt + M ( r) \ell d\theta,
\end{gather}
then the spin connections would be 
\begin{gather}
\omega_t^{01}=-\omega _t^{10} =- M', \ \ \ \ \  \ \ \ \omega_r^{02}=-\omega_r^{20}=-\sqrt{D} M' ,\nonumber\\
\omega_\theta^{01}=-\omega_\theta^{10} = MM'-N', \ \ \ \ \ \omega_\theta^{12}=-\omega_\theta^{21}=-\frac{M'}{2\sqrt{D}}.
\end{gather}
Then calculating different terms in the Lagrangian, we find
\begin{gather}
\epsilon_{ABC} R^{AB} e^C= \frac{3}{2} \ell M'^2 dt \ dr \ d\theta, \ \ \ \ \ \ \ \ \frac{1}{3 \ell^2} \epsilon_{ABC} e^A e^B e^C=-\ell dt \ dr \ d\theta, \nonumber\\
L_{\text{CS}}=2 \left(M' N''-M'' N' \right) dt \ dr \ d\theta, \ \ \ \ \ \ \ \ 2\lambda_A T^A=0.
\end{gather}
Taking the integral we get
\begin{gather}
\delta I_{\mathcal{M}}= -\frac{\ell}{8G} (\nu^2-1) (r_+-r_-) \delta t.
\end{gather}

From the boundary term we also get
\begin{gather}
\delta I _{\partial M}= \frac{\ell}{16G} (\nu^2+3)(r_+ -r_-) \delta t.
\end{gather}
So the rate of complexity growth would be
\begin{gather}\label{eq:warpgrowthTMG}
\dot{\mathcal{C}}= \frac{\ell}{G} \left(\frac{5-\nu^2}{16}\right) (r_+-r_-).
\end{gather}

As the central charges of dual CFT are
\begin{gather}
c_L=\frac{4 \nu \ell}{(3-\nu^2) G}, \ \ \ \ \ \ c_R=\frac{(5\nu^2-3) \ell}{\nu (3-\nu^2) G},
\end{gather}
 in order to have positive central charges we should have $\nu^2 <3$, then one can see the growth of complexity is actually positive as one expected.
 
It can be seen that the deformation parameter $\nu$ would actually decrease the rate of the growth of complexity. In future works, different proposed pictures for complexity, such as the ones in \cite{Freedman:2016zud} or \cite{Caputa:2017yrh} could be implemented to describe this fact, which we will explain them in the discussion section.

 The thermodynamical properties of warped $\text{AdS}_3$ black holes are also as follows \cite{Anninos:2008fx}
 
 \begin{flalign}
 T_R&=\frac{(\nu^2+3)(r_+-r_-)}{8 \pi \ell},\ \ \ \ \ \ 
 T_L=\frac{(\nu^2+3)}{8 \pi \ell} \left( r_++r_- - \frac{\sqrt{(\nu^2+3)r_+ r_-} }{\nu} \right),\nonumber\\
 T_H&=\frac{(\nu^2+3)}{4\pi \ell} \frac{(r_+-r_-)}{(2\nu r_+-\sqrt{(\nu^2+3)r_+ r_-} )},\ \ \ \ \ \ 
  S   = \frac{\pi \ell}{24 \nu G} \left [ (9\nu^2+3) r_+ - (\nu^2+3)r_-  - 4\nu \sqrt{(\nu^2+3)r_+ r_-} \right ].
 \end{flalign}
One can see that again the rate of complexity growth is more correlated with the temperatures, i.e., $T_R$ and $T_H$, rather than the entropy of the black hole. It could be interesting to try to further explain this observation by considering the properties and dynamics of the modes in the warped CFTs and then also by taking into account some other more exotic pictures such as $\text{ER}=\text{EPR}$ in warped CFTs or others such as \cite{Freedman:2016zud, Caputa:2017yrh}.
 
\subsection{Null Warped $\text{AdS}_3$}

A vacuum solution of TMG is null warped $\text{AdS}_3$ which is only well-defined at $\nu=1$. So it would be easier to study just the effect of $\mu$ term in the action on the rate of complexity growth. Its isometry group is again $SL(2,R) \times U(1)_{\null}$. Also the entropy and $T_L$ for this case is zero, but $T_R=\frac{\alpha}{\pi l}$. 

The metric of null warped black hole is of the form \cite{Chen:2009hg},
\begin{flalign}
ds^2 &= l^2 \left(-2r d\theta dt+(r^2+r+\alpha^2) +\frac{dr^2}{4r^2}    \right),
\end{flalign}
where to avoid naked causal singularity, one should take $0<\alpha <1/2$. 

The vierbein are
\begin{gather}
e^0  =r l dt+\frac{1}{2} l (1-\alpha^2-r-r^2) d\theta, \ \ \ \ \ 
e^1 = \frac{l}{2r} dr, \ \ \ \ \ 
e^2 =-r l dt+\frac{1}{2} l (1+\alpha^2+r+ r^2) d\theta,
\end{gather}
and the non-zero components of the spin connections are
\begin{flalign}
\omega^{01}&=-\omega^{10} = r dt+ \frac{1}{2} (1-r-3r^2+\alpha^2) d\theta, \nonumber\\
\omega^{02}&=-\omega^{20}=\frac{1}{2r} dr, \nonumber\\
\omega^{12}&=-\omega^{21}= r dt+\frac{1}{2} ( -1-r-3r^2 +\alpha^2) d\theta.
\end{flalign}

Now computing all the terms in the Lagrangian we get
\begin{gather}
\epsilon_{ABC} R^{AB} e^C= -3 l \ dt dr d\theta , \ \ \ \ \ \ \
\frac{1}{3 l^2} \epsilon_{ABC} e^A e^B e^C = l dt dr d\theta,  \nonumber\\
L_{CS}=2(1+\alpha^2+3r^2) dt d\theta dr,\ \ \ \ \ \ \ \  \ B=-4r l d\theta dt.
\end{gather}

Taking the integral from $r=0$ to a specific $r_s$ we will get
\begin{gather}
\dot{\mathcal{C}}=\frac{r_s}{4G} \left(l+\frac{1+\alpha^2+r_s^2}{2\mu} \right)-8\pi G r_s.
\end{gather}

The first two terms come from the bulk action and the last term comes from the boundary term. Note that again decreasing the parameter $\mu$, which increases the chirality, would actually increase the rate of growth of complexity similar to the BTZ solution of TMG.

\subsection{Supersymmetric black hole}

A new solution of TMG with negative cosmological constant were found in \cite{Dereli:2000fm} which is supersymmetric, asymptotically approaches the extremal BTZ solution, and goes to flat space if one sets the cosmological constant to zero. So with these specific characteristics it might be interesting to also check its rate of complexity growth. 

For these black holes the vierbein would be \cite{Dereli:2000fm},
\begin{flalign}
e^0= f(\rho) dt, \ \ \ \ \ \ \ e^1=d\rho,  \ \ \ \ \ \ \ \ e^2=h(\rho) (d\phi+a(\rho) dt),
\end{flalign}
and the spin connections are
\begin{flalign}
{\omega^0}_1=\left(f'-\frac{a a' h^2}{2f}\right)dt-\frac{a' h^2}{2f}d\phi, \ \ \ \ \ \ {\omega^0}_2=-\frac{a' h}{2f} d\rho, \ \ \ \ \ {\omega^1}_2=\left( -\frac{a' h}{2}-ah'\right)dt -h' d\phi.
\end{flalign}

The metric functions for the solution of  \cite{Dereli:2000fm}   are
\begin{flalign}
f&=f_0 e^{2\rho/l} \left(1+ \beta_1 e^{2\rho/l} +\beta_2 e^{(1/l-k \mu) \rho} \right)^{-1/2}, \nonumber\\
h&=h_0 \left( 1+\beta_1 e^{2\rho/l}+\beta_2 e^{(1/l-k \mu)\rho} \right)^{1/2}, \nonumber\\
a&=-a_0+k \frac{f_0}{h_0} e^{2\rho/l} \left(1+\beta_1 e^{2\rho/l}+\beta_2 e^{(1/l-k \mu) \rho} \right)^{-1},
\end{flalign}
where $\beta_1$, $\beta_2$, $a_0$, $f_0$, $h_0$ are some integration constants. The extremal BTZ
can be recovered in the limit of $ \big | \mu \big | \to \infty$ of the above solution. 

Note that both the extremal BTZ and the solution here are in fact supersymmetric since for them, there exist a 2-spinor $\epsilon$ which satisfies
\begin{gather}
(2\mathcal{D}+\frac{1}{l} \gamma) \epsilon=0,
\end{gather} 
where $\gamma=\gamma_a e^a$ and $\mathcal{D}=d+\frac{1}{2} \omega^{ab} \sigma_{ab}$.

As mentioned in \cite{Dereli:2000fm}, depending on the values of the integration constants $\beta_1$ and $\beta_2$, we can find singularities and event horizons in the metric functions. 

Now by inverting the metric and finding the roots of the $tt$ component of the inverse metric, i.e,  $g^{-1}_{tt}$, we can find the location of the event horizon as
\begin{gather}
2 \beta_1 e^{\left(\frac{1}{l}+k \mu \right) \rho }+\beta_2(1-\mu k l ) =0,  \ \ \ \to \  \rho _s=\frac{l}{1+\mu k l  }\log \left(\frac{\beta_2 (\mu k l -1) }{2 \beta_1}   \right).
\end{gather}

Now for this solution we can find the terms in the Lagrangian.  The first term gives
\begin{flalign}
\epsilon_{ABC} R^{AB} e^C&= \frac{a'^2 h^3}{2f} -2( h f''+h' f' + h'' f ),
\end{flalign}
The second term would be
\begin{flalign}
\frac{1}{3 l^2} \epsilon_{ABC} e^A e^B e^C=\frac{2 f h}{l^2} dt d\phi d\rho.
\end{flalign}
The Chern-Simons term is
\begin{flalign}
L_{CS}&= \frac{h^2 a'}{f^2} \left( f'^2-h^2 a'^2 \right)- \frac{f' h}{f} \left( a'' h+4 a' h' \right)
+h \left( a'' h'-a' h''+\frac{a' h f''}{f} \right) + 3 a' h'^2,
\end{flalign}
and the boundary term is
\begin{flalign}
B=-\frac{1}{16\pi G} \left( h f'+h' f \right) d\phi \wedge dt= -\frac{1}{4 G l} e^{\frac{2 \rho }{l}} f_0 h_0 \delta t.
\end{flalign}

The general terms for the rate complexity is complicated. However for the special case of $k=1$ in this solution, things would become much more simplified and therefore we only present the result for this special case. Therefore we find,
\begin{flalign}
\delta I_M&= \frac{f_0 h_0 }{ G l^2}\left(\left(\frac{\beta _2 (\mu  l-1)}{2\beta _1}\right)^{\frac{2}{\mu  l+1}}-\frac{l}{4}\right) \delta t,
\end{flalign}
and
\begin{flalign}
\delta I_{\partial \mathcal{M}}=\frac{f_0 h_0}{G l} \left(\frac{1}{4}-2^{-2-\frac{2}{ \mu  l+1}} \left(\frac{\beta _2(\mu  l-1)}{\beta _1}\right){}^{\frac{2}{ \mu  l+1}}\right)\text{$\delta $t},
\end{flalign}
So the rate of complexity growth would be
\begin{gather}
\mathcal{\dot{C}}=\frac{ f_0 h_0 }{G l^2}2^{-2-\frac{2}{ \mu  l+1}}\left(4^{\frac{1}{ \mu  l+1}} l-(l-2) \left(\frac{\beta _2 (\mu  l-1)}{\beta _1}\right){}^{\frac{2}{\mu  l+1}}\right).
\end{gather}

Note that it is actually a more complicated function of $\mu$, but generally it decreases by increasing the coupling constant $\mu$ similar or BTZ and warped $\text{AdS}_3$ case.

\subsection{ACL black hole}

In addition to BTZ, TMG also admits a non-asymptotically flat, non-asymptotically AdS black hole solution named ACL \cite{Bouchareb:2007yx}. It was shown in \cite{Moussa:2008sj} that these black holes are geodesically complete and causally regular which this property makes the computation of their complexity interesting. 

 This black hole is of the following form
\begin{flalign}
ds^2&=-\beta^2 \frac{\rho^2-\rho_0^2}{r^2} dt^2+\frac{1}{\zeta^2 \beta^2} \frac{d\rho^2}{\rho^2-\rho_0^2}+r^2 \left( d\varphi-\frac{\rho+(1-\beta^2)\omega}{r^2} dt \right)^2,
\end{flalign} 
with
\begin{flalign}
r^2= \rho^2+2 \omega \rho+\omega^2 (1-\beta^2)+\frac{\beta^2 \rho_0^2}{1-\beta^2},
\end{flalign}
where
\begin{gather}\label{eq:change}
\beta^2 \equiv \frac{1}{4} \left( 1-\frac{27 \Lambda}{\mu^2} \right), \ \ \ \ \ \zeta=\frac{2}{3}\mu.
\end{gather}

Note that the two parameters $\omega$ and $\rho_0 \ge 0$ are related to the mass and angular momentum of the black hole. 
Also if $\omega=\rho_0=0$, the metric becomes horizonless and becomes a ground state solution. Therefore, we expect for this case the complexity and its growth rate vanishes.

Writing the metric in the ADM form
\begin{gather}
ds^2=-N^2 dt^2+r^2 (d\varphi+N^{\varphi} dt)^2+\frac{1}{(\zeta r N)^2} d\rho^2,
\end{gather}
the dreibein $e^a$ for this metric would be
\begin{gather}
e^0=N dt,  \ \ \ \ \ \ \ \ \ e^1= r( d\varphi+N^{\varphi} dt), \ \ \ \ \ \ \ \ e^2=\frac{1}{\zeta r N} d\rho, 
\end{gather}
with the following corresponding spin connections \cite{Bouchareb:2007yx}, 
\begin{flalign}
{\omega^0}_2 &=\zeta r [N' e^0+\frac{1}{2} r (N^\varphi)' e^1],\nonumber\\
{\omega^0}_1&=  \zeta r \frac{1}{2} r (N^\varphi)' e^2, \nonumber\\
{\omega^1}_2& = \zeta r [\frac{1}{2} r (N^\varphi)' e^0+N \frac{r'}{r} e^1].
\end{flalign}
Now calculating different terms in the Lagrangian, we get the following results
\begin{gather}
\epsilon_{ABC} R^{AB} e^C=\frac{\zeta r}{2} \left (r^3 ({N^\varphi}')^2+4 N N' r' \right)dt d\rho d\varphi= \frac{\zeta}{2}dt d\rho d\varphi,  \nonumber\\
 \frac{1}{3 l^2}\ \epsilon_{ABC}  e^A e^B e^C =-\frac{2}{\zeta l^2} dt d\rho d\varphi,  \nonumber\\
L_{CS}=-\frac{\zeta^2 r^3}{2}  (N^{\varphi})'   \left(r^3 ({N^\varphi} ')^2 - 4 N N' r' \right)dt d\rho d \varphi,
\end{gather}
 and the boundary term would be
 \begin{gather}
 \frac{1}{32 \pi G} \epsilon_{ABC} \omega^{AB} e^C= \frac{\zeta }{16 \pi G} r N ( N' r+N r') dt d\varphi.
 \end{gather}
 
The contribution of the first two Einstein terms would be
\begin{gather}
\delta  I_{\mathcal{M}_1}=-\frac{\rho_0 \zeta  }{8 G} \left(1-\frac{4}{l^2 \zeta^2 } \right) \delta t,
\end{gather}
and the contribution of the Chern-Simons term is
\begin{gather}
\delta I_{\mathcal{M}_{CS}}=\frac{\zeta  \beta }{24 G } \Bigg (\frac{\zeta ^2\left(2 \rho _0{}^2+\omega ^2\left(3 \beta ^4-\beta ^2-2\right)\right)}{\sqrt{\left(\rho _0{}^2+\left(\beta ^2-1\right) \omega ^2\right)\left(\beta ^2-1\right)}}\text{ArcTanh}\left(\frac{2 \beta \text{  }\rho _0 \sqrt{\left(\rho _0{}^2+\left(\beta ^2-1\right) \omega ^2\right)\left(\beta ^2-1\right)}}{\rho _0{}^2+\left(\beta ^4-1\right) \omega ^2}\right)\nonumber\\
+5\beta  \omega  \log \left(\frac{\rho _0-\omega \left(\beta ^2-1\right) }{\rho _0+\omega \left(\beta ^2-1\right) }\right)+\frac{4 \rho _0{}^3\beta ^3 }{\rho _0{}^2-\left(\beta ^2-1\right)^2 \omega ^2}-6\rho _0 \beta -\frac{\rho _0}{\beta }\Bigg) \delta t.
\end{gather}
Finally the boundary term would result in
\begin{gather}
\delta I_{\partial \mathcal{M}}= \frac{ \rho _0 \zeta \beta ^2 }{4 G}\text{$\delta $t}.
\end{gather}
\begin{figure}[ht!]
 \centering
  \includegraphics[width=75mm]  {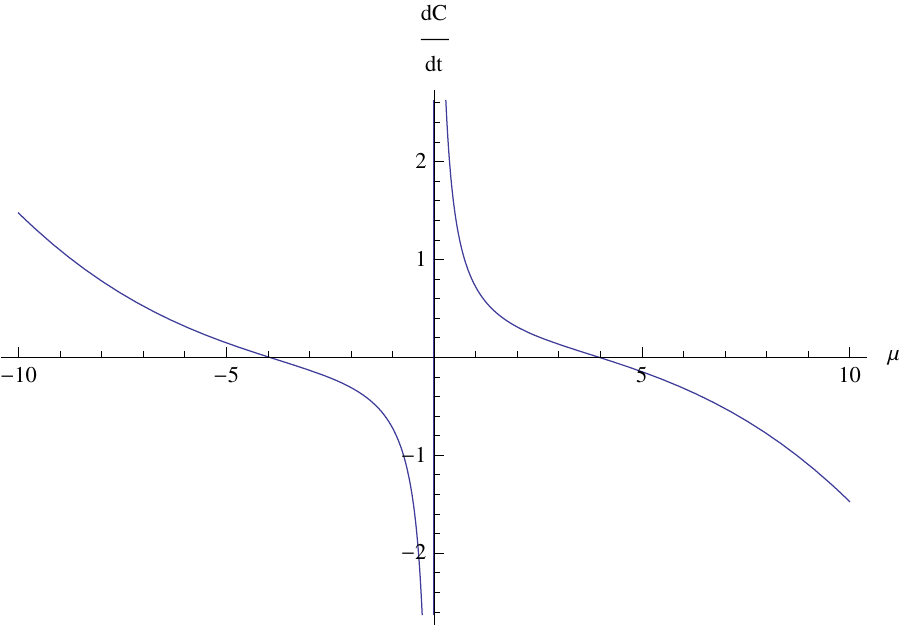}
  \caption{Plot of $\dot{\mathcal{C}}$ vs. $\mu$ for $\rho_0=\omega=G=l=1$ and $\beta=\frac{1}{2}$.  }
  \label{fig:ACL}
\end{figure}
Note that if $\omega=\rho_0=0$ all these three terms vanish as we have expected. 

Using \eqref{eq:change}, one can then write the sum of all these in terms of $\mu$. The final result of the rate of complexity growth versus $\mu$ is shown in Figure \ref{fig:ACL}. Again, one can notice that decreasing $\mu$, meaning increasing the effect of Chern-Simons term, would increase the rate of complexity growth to the point that for $\mu \to 0$ it diverges, again in a $\frac{1}{\mu}$ fashion, which is similar to the previous cases. For this figure, however, note that one should only consider the region where $\dot{\mathcal{C}}$ is positive.

As calculated in \cite{Bouchareb:2007yx}, the thermodynamical quantities of this black hole is as follows
\begin{gather}
T_H=\frac{\mu \beta^2}{3\pi} \frac{\rho_0 \sqrt{1-\beta^2} }{ \rho_0+(1-\beta^2) \omega}, \ \ \ \ \ \ 
S=\frac{\pi}{3G \sqrt{1-\beta^2} } \left( (1+\beta^2)\rho_0+(1-\beta^2)\omega \right),\nonumber\\
\mathcal{M}=\frac{\mu}{9 G} \beta^2 (1-\beta^2)\omega, \ \ \ \ \ \ \ \mathcal{J}=\frac{\mu \beta^2 }{18G} \left( (1-\beta^2) \omega^2-\frac{1+\beta^2}{1-\beta^2}\rho_0^2 \right). 
\end{gather}

Here it is a bit more difficult to distinguish the strength of correlation between different thermodynamical quantities.

\subsection{Shockwave solution of TMG}

Similar to \cite{Alishahiha:2017hwg}, one can also study the shockwave solution of TMG to get more information about the boundary complexity. We will present this computation in our future paper \cite{Ghodrati:2017rta}. However we just sketch the general idea here.

First, one writes the black brane metric of
\begin{gather}
ds^2=-\frac{r^2-r_h^2}{\ell^2} dt^2+\frac{\ell^2}{r^2-r_h^2} dr^2+\frac{r^2}{\ell^2} dx^2, \ \ \ \ \ \ \ \ \ \ \ \ \ \ \Lambda=-\frac{1}{\ell^2},
\end{gather}
in the Kruskal coordinates, \cite{Alishahiha:2016cjk}
\begin{gather}
ds^2= 2 A(uv) du dv+B(uv) dx^2,
\end{gather}
where
\begin{gather}
A(uv)=-\frac{2 c \ell^2}{(1+ c uv)^2}, \ \ \ \ \ \ \ \ B(uv)=\frac{r_h^2}{\ell^2} \left( \frac{1-c u v}{1+c u v} \right).
\end{gather}
Then, by considering the following back-reacted metric ansatz
\begin{gather}
ds^2= 2 A(UV) dU dV+B(UV) dx^2 -2 A(UV) h(x) \delta(U) dU^2,
\end{gather}
and the calculated form of shock wave strength, i.e, the function $h(x)$, \cite{Alishahiha:2016cjk}
\begin{gather}
h(x)=-\frac{\eta}{2 a_2} \left (x+\frac{1}{2 a_2} \right) e^{-a_2 x},
\end{gather}
and also the following form of scrambling time and butterfly velocities \cite{Alishahiha:2016cjk}
\begin{gather}
t_*=-\frac{\beta}{2\pi} \log \frac{k}{\ell},  \ \ \ \ \ \ \ \ \ \ \ \ \ v_B^{(1)}=\frac{2\pi}{\beta a_2}=1,  \ \ \ \ \ \ \ \ \ \ \ \ \ v_B^{(2)}=\frac{2\pi}{\beta a_1}=\frac{1}{\mu \ell},
\end{gather}
one can calculate the action for two different regimes of small shifts, $ u_0^{-1} +h(x) <v_0$ and large shifts where $ u_0^{-1} +h(x) \ge v_0$. Note that for these two different cases the corresponding WDW patches are different leading to different results for the rate of complexity growth.

\section{Complexity growth in a parity-preserving theory}

We now study the rate of complexity growth for several black hole solutions of New Massive Gravity, as a parity-preserving theory, to compare some results with the previous case and also to examine the effects of mass term, warping factor, hair parameter and different variables of the theory. 

The action of NMG is 
\begin{flalign}\label{eq:NMGaction}
I=\frac{1}{16 \pi G} \int_{\mathcal{M}} d^3 x \sqrt{-g} \left[ R-2\Lambda-\frac{1}{m^2} \left( R^{\mu \nu} R_{\mu \nu} -\frac{3}{8} R^2 \right) \right],
\end{flalign}
where $m$ is a dimensionful parameter. 
One can also write this theory in the following form \cite{Hohm:2010jc}
\begin{gather}
I=\frac{1}{16 \pi G} \int_{\mathcal{M}} d^3 x \sqrt{-g} \left[ R-2\Lambda+f^{\mu \nu} G_{\mu \nu}+\frac{m^2}{4} \left( f^{\mu \nu} f_{\mu \nu} -f^2 \right) \right ].
\end{gather}
In the above term, $G_{\mu \nu}$ is the Einstein tensor and the auxiliary field $f_{\mu \nu}$ is
\begin{gather}
f_{\mu \nu}=-\frac{2}{m^2} \left( R_{\mu \nu}-\frac{1}{2 (d+1)} R g_{\mu \nu} \right ).
\end{gather} 
The Gibbons-Hawking boundary term would be
\begin{gather}\label{eq:NMGboundary}
I_{GGH}= \frac{1}{16 \pi G} \int_{\partial \mathcal{M}} d^2 x \sqrt{- \gamma} \left( -2K-\hat{f}^{ij} K_{ij}+\hat{f} K \right),
\end{gather}
where $K_{ij}$ is the extrinsic curvature of the boundary and $K=\gamma^{ij} K_{ij}$ is the trace of the extrinsic curvature.  The auxiliary filed $\hat{f}^{ij}$ is also defined as
\begin{gather}
\hat{f}^{ij}= f^{ij}+2h^{(i} N^{j)}+s N^i N^j,
\end{gather}
where above functions are defined from the following ADM form of the metric
\begin{gather}
ds^2= N^2 dr^2+\gamma_{ij}( dx^i+N^i dr) (dx^j+N^j dr).
\end{gather}

Note that NMG is also a rich theory which admits several solutions. The complexity growths for BTZ, AdS-Schwarzschild black hole, and shockwave solutions of this theory have been studied in \cite{Alishahiha:2017hwg}.  Here we would like to study the rate of complexity for some other black hole solutions.

\subsection{Warped $\text{AdS}_3$ black hole}
The form of the metric has been given in section \eqref{eq:warped}. By calculating the action \eqref{eq:NMGaction} and the boundary term \eqref{eq:NMGboundary}, the rate of complexity growth can be found as
\begin{flalign}
\delta I_{\mathcal{M}}&= \frac{1}{16 \pi G} \int_{r_-}^{r_+} \int_{t}^{t+\delta t} \int_0^{2\pi} \mathcal{L}_{\text{NMG}} d\phi dt dr\nonumber\\&
=\frac{\delta t \ l \left(4 \nu ^4-48 \nu ^2+9\right)}{8G(20 \nu ^2-3)} (r_+-r_-),
\end{flalign}
and 
\begin{flalign}
\delta I_{\mathcal{\partial M}}&= \frac{1}{16 \pi G} \int_t^{t+\delta t} \int_0^{2\pi} dt \ d\phi \left[ \frac{3 l\left(\nu ^2+3\right)\left(4 \nu ^2-1\right)}{ \left(20 \nu ^2-3\right)}\left(2 r-r_+-r_-\right)\right] \Bigg | _{r+}\nonumber\\& - \frac{1}{16 \pi G} \int_t^{t+\delta t} \int_0^{2\pi} dt \ d\phi \left[ \frac{3 l\left(\nu ^2+3\right)\left(4 \nu ^2-1\right)}{ \left(20 \nu ^2-3\right)}\left(2 r-r_+-r_-\right)\right] \Bigg | _{r-}
\nonumber\\& = \frac{\delta t \  3 l (\nu^2+3) (4\nu^2-1)}{4G (20\nu^2-3)} (r_+-r_-).
\end{flalign}
So, the rate of increase in complexity is
\begin{gather}\label{eq:warpgrowthNMG}
\dot{\mathcal{C}}=\frac{dI}{dt} =\frac{l \ \left(28 \nu ^4+18 \nu ^2-9\right)}{8 G \left(20 \nu ^2-3\right)}\left(r_+-r_-\right).
\end{gather}
We can also write the above result in terms of the conserved charges of the solution, $\mathcal{M}$ and $\mathcal{J}$ \cite{Ghodrati:2016vvf}.

One can notice that similar to \eqref{eq:warpgrowthTMG}, both in TMG and NMG, the rate of complexity growth for warped $\text{AdS}_3$ black holes only depends on the warping factor $\nu$ and the difference between the inner and outer horizons; i.e. $\dot{\mathcal{C}}_\text{WBTZ} \propto (r_+-r_-)$, while in the BTZ case the relation is in the form of $\dot{\mathcal{C}}_{\text{BTZ}} \propto (r_+^2-r_-^2 )$. These results are summarized in Table \ref{tab:table1}.  \\ 

\begin{table}[h]
\centering
\begin{tabular}{l | l | l} 
 & \ \ \ \ \ \  TMG &  \ \ \ \ \ \  NMG \\ 
\hline  \rule{0pt}{1.5\normalbaselineskip}
BTZ & $\frac{r_+^2- r_-^2}{4 G l^2} \left (\frac{r_+^2 +r_-^2}{\mu l   r_+ r_-} \right)$ & $\frac{r_+^2- r_-^2}{4 G l^2} \left (1-\frac{1}{2 l^2 m^2} \right)$ \\ [2.5ex]
$\text{WAdS}_3$ & $\frac{l \left(r_+-r_-\right)}{8G}\left( \frac{ 5-\nu^2}{2}\right)$
 & $\frac{l \left(r_+-r_-\right)}{8G}\left( \frac{ 28 \nu ^4+18 \nu ^2-9}{20 \nu ^2-3}\right) $
\end{tabular}
\caption{Complexity growth of BTZ and $\text{WAdS}_3$ black holes in two theories of TMG and NMG.}
\label{tab:table1}
\end{table}

Interestingly, this is similar to the way that the inner and outer temperatures of the horizons of these black holes, i.e, ${T_H}^+$, ${T_H}^-$,  and also the right-moving temperature $T_R$ of warped $\text{AdS}_3$ black hole depend on the horizons' radii. The difference between the factors of $r_+$ and $r_-$ in CFT and warped CFT cases could be due to the fact that In CFTs we have both left and right moving modes while in WCFTs there are only right moving modes. Studying these relations further could help to a better understanding of the thermodynamics of quantum complexity. 

 Also, it is worth to notice that in the region where the solution is spacelike stretched and is free of naked Closed Timelike Curves (CTCs), (which is satisfied when $\nu^2 >1$), the relation \eqref{eq:warpgrowthNMG} is an increasing function of $\nu$, while relation \eqref{eq:warpgrowthTMG} is a decreasing one.

\subsection{New hairy black hole}
It would also be interesting to study the effect of black hole's hair on the growth rate of complexity as any hair parameter could change different features of black holes such as evaporation, encoding of information and scrambling behaviors. 

For doing so we study a hairy black hole solution of NMG which was first introduced in \cite{Oliva:2009ip} and then later it was studied more in \cite{Nam:2010dd, Giribet:2009qz , Capela:2012uk} and also in \cite{Ghodrati:2016ggy, Ghodrati:2016tdy} where their Hawking-Page phase diagrams were presented.

For this type of black hole, in the action \eqref{eq:NMGaction}, we should set $m^2=\Lambda=-\frac{1}{2 l^2}$. Then the form of the metric could be derived as
\begin{gather}
ds^2=-N F dt^2+\frac{dr^2}{F}+r^2 (d\phi+N^\phi dt)^2,
\end{gather}
where
\begin{equation}
\begin{split}
N=&\Big( 1+\frac{b l^2}{4H} \big( 1-\Xi^{\frac{1}{2}} \big) \Big)^2,\ \ \ \ \ \ \ \ \ \  
N^\phi=-\frac{a}{2r^2} (4G_N M-bH),\nonumber\\
F=&\frac{H^2}{r^2} \Big( \frac{H^2}{l^2} +\frac{b}{2} \Big(1+ \Xi^{\frac{1}{2}} \Big)H+\frac{b^2 l^2}{16} \Big( 1-\Xi^ {\frac{1}{2}} \Big)^2- 4G_N M\ \Xi^{\frac{1}{2}} \Big),\nonumber\\
H=& \Big( r^2-2G_N M l^2 \big (1- \Xi^{\frac{1}{2}} \big) -\frac{b^2 l^4}{16} \big(1-\Xi^{\frac{1}{2}} \big)^2 \Big)^{\frac{1}{2}},
\end{split}
\end{equation}
and $\Xi :=1-\frac{a^2}{l^2}$ , $-l \le a \le l $. 

Depending on the range of the parameters $M$, $a$, and $b$, this solution could have an ergosphere and inner and outer horizons which could make this example more interesting for studying its rate of complexity growth.

The Penrose diagrams for different signs of $b$ and $\mu$ have been brought in \cite{Oliva:2009ip} which basically are similar to Schwarzschild case. 

Now calculating the complexity growth for the most general case gives a very complicated answer. Since we are only interested in studying the effect of hair parameter $b$ here, by taking $a=0$, we only consider the non-rotating case. So we get
\begin{flalign}
\delta I_{\mathcal{M}}&=I_{\mathcal{M}} [\text{WDW}  |_{t+\delta t}]- I_{\mathcal{M}} [\text{WDW}  |_t]=\frac{\left(r_+-r_-\right)}{2 G l^2} \left(b l^2+r_++r_-\right)\text{$\delta $t},
\end{flalign}
and the contribution from the boundary term is
\begin{flalign}
\delta I_{\partial \mathcal{M}} =-\frac{\left(r_+-r_-\right)}{2 G l^2}\left( \frac{b^2 l^4}{4}+b l^2 \left(r_++r_-\right)+\frac{2 }{3}\left(r_+{}^2+r_+ r_-+r_-{}^2-6 M G l^2\right)\right)\text{$\delta $t}.
\end{flalign}
Note that increasing the hair parameter $b$ increases the contribution to the complexity growth  from the bulk term and decreases the complexity growth coming from the boundary term.

For the following special case where 
\begin{gather}
b=-\frac{1}{l^2} (r_++r_-), \ \ \ \ \ \ \ M=-\frac{r_+ r_-}{4 G l^2},
\end{gather}
the total rate of complexity growth is
\begin{gather}
\dot{\mathcal{C}}=\frac{\left(r_+-r_-\right) }{2 G l^2}\left(\frac{1}{3}\left(r_+^2+r_-^2\right) + \frac{l^2}{4}\left(r_++r_-\right) +\frac{7}{3} r_+r_-\right).
\end{gather}
One can see that similar to the relation for the temperature of this type of black hole, the rate of complexity growth also has a factor of $(r_+-r_-)$.

It might also be interesting to study complexity growth rate for the case with a positive cosmological constant where the black hole posses the event and cosmological horizons \cite{Oliva:2009ip}. 

\subsection{Log black hole solution } 
Another solution of NMG which one might think that the behavior of its complexity growth could be interesting is the so called ``log" solution. This solutions was first found in \cite{Clement:2009ka,Ghodsi:2010gk}. For the special case of $\nu=-l$ which were defined in \cite{Ghodsi:2010gk}, the following simple solution could be written as
\begin{flalign}
ds^2&=\left ( -2 l^{-2} \rho+( A \rho^{-l} \log (\rho) +B) \right) dt^2-2l ( A \rho^{-l} \log(\rho)+B) dt d\phi\nonumber\\ &+
\left( 2\rho +l^2 (A \rho^{-l} \log(\rho) +B)\right) d\phi^2+\frac{l^2 d\rho^2}{4n\rho^2}, 
\end{flalign}
and for this case the entropy is
\begin{gather}
S=\frac{2 A_h l (l+1)}{G (1+8l(l+1)}.
\end{gather}

Now computing the Lagrangian \eqref{eq:NMGaction} and \eqref{eq:NMGboundary} we find
\begin{gather}
\delta I_{\mathcal{M}}=-\frac{4 (l+1) \left(r_+-r_-\right)}{G l (1+8 l (l+1))}\text{$\delta $t},
\end{gather}
and
\begin{gather}
\delta I_{\partial \mathcal{M}}=\frac{4(l+1) \left({r_+}^2-{r_-}^2\right)}{G l (1+8 l (l+1))}\text{$\delta $t},
\end{gather}
leading to the total complexity growth of
\begin{gather}
\dot{\mathcal{C}}=\frac{4 (l+1)}{G l (1+8 l (l+1))}  \left(r_+-r_-\right) \left(r_++r_--1\right).
\end{gather}

One can see that even for a ``log" solution, although different terms such as $f_{ij}$ might have a complicated form, but the final result for the rate of complexity growth would greatly simplify and also a factor of $(r_+ - r_-)$ is again present here, similar to the BTZ and warped $\text{AdS}_3$ black holes.

\section{Discussion}\label{sec:disc}

In this paper our first aim was to examine the effect of chirality on the rate of growth of complexity in order to get more information about how different aspects of ``Complexity=Action" conjecture would work in different setups. 

To do so we studied the rate of complexity growth for different solutions of a chiral breaking theory (TMG) and a chiral-preserving theory (NMG). Specifically using CA conjecture, we calculated the complexity growth for BTZ, warped $\text{AdS}_3$, null warped $\text{AdS}_3$, supersymmetric and ACL black hole solutions of TMG and then warped $\text{AdS}_3$, new hairy and log black hole solutions of NMG.

Using the specific Gibbons-Hawking boundary term of TMG, introduced in \cite{Miskovic:2009kr}, and then by calculating different terms of the action and integrating on the Willer-DeWitt patch, we found that increasing the parameter $\mu$ would actually decrease the rate of complexity growth of BTZ black hole in TMG. By decreasing $\mu$ which increases the effect of Chern-Simons term and increases chirality, the rate of complexity growth would increase where for $\mu \to 0$ the rate of complexity growth would diverge. For the parity-preserving theory of NMG, however, we see that by decreasing $m^2$ which increases the effect of higher derivate term (couples the $\frac{1}{m^2}$),  the rate of growth of complexity of BTZ would decrease. 

For the case of warped $\text{AdS}_3$ black hole we found that generally the warping factor $\nu$ decreases the rate of growth of complexity in the chiral theory of TMG while it increases this rate in  NMG. This could be interpreted by the dynamics of left and right moving modes and their effects on the growth of complexities in these two theories. It would also be very interesting to study the effect of $\mu$ or warping factors on the scrambling or switchback time of a warped $\text{AdS}_3$ black hole in these two theories and compare the results.

Another interesting point that we have found is that in all of these theories there was a direct relationship between the rate of complexity growth and the difference between the inner and outer horizons; i.e, $\dot{\mathcal{C}} \propto (r_+-r_-) $. This factor is also present in the relation of temperature of BTZ, Warped BTZ and hairy black holes while this is not the case for the relations of entropies. This could suggest that there is a stronger correlation between the complexity and temperature, rather than the complexity and entropy, which is worth further investigation. This is actually in line with the idea of Lloyd \cite{Lloyd:2000}.  This fact could help to understand further the similar thermodynamical laws for complexity. 

One could also think to study this rate for other solutions such as Lifshitz or hyperscaling violating \cite{Ghodrati:2014spa} backgrounds with black holes. However for those solutions which breaks the Lorentz invariance there is not a well-defined Carter-Penrose diagram as the scaling of time and space coordinates for these backgrounds are different and this makes the form of WDW patch and computing the holographic complexity more difficult. 

It would also be very interesting to calculate the rate of change of complexity in the dynamical setups such as the ones in \cite{Sachs:2011xa, Flory:2013ssa} where an interpolating solution between a past horizon and a chiral AdS pp-wave has been found.  Studying the behavior of complexity growth rate and the effects of different factors in these backgrounds could shed more light on the nature of holographic complexity.

In \cite{Carmi:2016wjl}, the structure of UV divergences that appear in  $\mathcal{C}_V (\Sigma )$ has been studied where for the first order the coefficients of the divergences have been written in terms of the extrinsic and intrinsic curvatures which were integrated over the boundary time slice. To gain more information about these structures, in different boundary CFTs such as warped CFTs or topological matters and to study the effect of different factors such as chirality, it might be useful to go beyond the first order and to try to find some universal coefficients in the first and second order of these divergences. For example similar to \cite{Alishahiha:2017cuk} one can gain more information about the fidelity susceptibility of different field theories. We will sketch few steps for this calculation in appendix \ref{appendix}.

 There are many more ideas and progresses in different setups for calculating complexity or complexity growth rate that could be applied for the warped CFT case as well, where in the following parts we are going to review.  

In \cite{Hashimoto:2017fga}, by discretizing the $U(1)$ gauge group as $\mathbf{Z}_N$ authors studied the time evolution of complexity for Abelian pure gauge theories. They could define a universal gate set for the $U(1)$ gauge theories which enabled them to calculate the complexity growth explicitly. It would be interesting to use the same idea for WCFT and by discretizing the $U(1)_L \times SL(2, \mathbb{R})_R$ group, define some new gate sets.  Then by evaluating the eigenvalues of the Hamiltonian, one could directly study the rate of complexity growth in WCFTs and then one can compare the results with usual CFTs and also obviously with the results found here coming from holography.

Another approach to define complexity was introduced by Nielsen,\cite{Nielsen:2005}. In this method, the complexity was defined by the geodesic distance between an unitary operator $U$ and an identity operator with respect to a metric in the space of unitary operators.  It would be interesting to study the behavior of complexity metric or Nielsen geometry in warped CFTs. Note that as mentioned in \cite{Brown:2017jil, Brown:2016wib}, the complexity metric would actually punish directions that touch more qubits. So it would be interesting to see how chirality and parity violation of the modes would affects the complexity metric. 

There are also some new ideas to evaluate complexity by minimizing the Liouville action, \cite{Caputa:2017yrh, Czech:2017ryf}
\begin{flalign}
S_L&=\frac{c}{24 \pi} \int dx \int_\epsilon^\infty dz \Big[ \underbrace{ (\partial_x \phi)^2+ (\partial_z \phi)^2}_{ \# \text{ of Isometries \includegraphics[width=4mm]{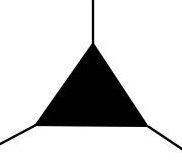} }} +\underbrace{\delta^{-2} e^{2\phi}}_{\# \text{ of Unitaries \includegraphics[width=4mm] {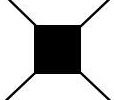} } }\Big].
\end{flalign}
By minimizing this action, one can define complexity \cite{Caputa:2017yrh} in the field theory side. It would also be possible to derive Einstein's equation in the bulk and to build a hyperbolic space which is the time-slice of $\text{AdS}_3$ \cite{Czech:2017ryf}. Note that in the Liouville action the first two terms which are the Kinetic terms are actually dual to the number of isometries (coarse-graining)  \cite{Czech:2017ryf} and the third term which is the potential term is dual to the number of unitaries (disentanglers), in the tensor network formalism of MERA.

It would be interesting to see if one can also derive the time slice of $\text{AdS}_3$ or warped $\text{AdS}_3$ space-times from warped CFTs by using the \textit{``chiral Liouville theory"} \cite{Compere:2013aya}, which is in the following form
\begin{flalign}
S_L&=\frac{c}{12 \pi} \int d^2 x \left(\partial_{+} \rho  \partial_{-} \rho -\frac{\Lambda}{8} e^{2\rho} + h(\partial_{-} \rho)^2 + [ \partial_- h \partial_- \rho ] -\frac{6}{c} h \Delta \right ),
\end{flalign}
or one can also write it as
\begin{flalign}
S&=S_L^0 +\int dt^+ dt^- \left( \underbrace{\partial_+ \phi \partial_- \phi}_{\# \text{  of Isometries}} +\color{blue}\underbrace{h \partial_- \phi \partial_- \phi}_{\text{$\#$ of WCFTs new gate}}\color{black}-\underbrace {\frac{m^2}{4} e^{2\rho} \phi^2 }_{\# \text{ of Unitaries}} \right) .
\end{flalign}

The main difference between the two actions is the middle term written in blue, i.e, $h \partial_- \phi \partial_- \phi$. As noted in \cite{Compere:2013aya}, $h$  which is proportional to a right moving current is dimension $(1,0)$ and $(\partial_- \phi)^2$ is dimension $(0,2)$. So this term is a dimension $(1,2)$ operator and one can think of chiral theory as a usual field theory which is deformed by this operator \cite{Compere:2013aya}. These operators are indeed very special as they are related to the $\text{IR}$ limits of the dipole deformed gauge theories. It would be interesting to study the corresponding gates for these operators in the MERA pictures and also study the effects of these operators on complexity or its rate of growth, and then compare these results with the holographic ones.

In \cite{Freedman:2016zud, Bao:2016rbj}, a new picture based on max-flow min-cut theorem or a set of Planck-thickness ``bit threads" for entanglement entropy has been proposed. In this picture entanglement entropy of a boundary region is related to the maximum number of threads that can emanate from its area.  One might be able to use this picture here as well and by considering the dynamics of these threads explains the holographic conjecture for the complexity and then use these ideas to explain the complexity of chiral theories explicitly.

In \cite{Bhattacharya:2014vja}, authors tried to define a quasi-local measure of quantum entanglement and by considering the infinitesimal variation of the region, they defined the concept of entanglement density. Also using the positivity of the entanglement which would be mapped to the null energy condition in the gravity bulk dual, they derived the second law of thermodynamics for the extremal surfaces. One might think that the similar concepts of quasi-local measure of quantum complexity and a notion of quantum complexity density could also be defined and using the positivity of complexity growth rate, similar maps to the bulk dual can be implemented which might lead to more rigorous thermodynamics-like laws for complexity.

Another more exotic idea is to study the relationship between complexity and Schiwnger effect. For doing so, similar to studies in \cite{Ghodrati:2015rta}, one could study both of their rates in different backgrounds and also study the effect of external factors such as electric or magnetic fields to find the relationship between these two quantities. Hypothetically, these studies could shed more light on the dynamics and structure of the bulk and also different aspects of $\text{ER}=\text{EPR}$.

\section*{Acknowledgement}
I would like to thank M. Alishahiha, S. Sadeghian, M. H. Vahidinia, A. Naseh, A. Faraji,  K. Hajian and H. Soltanpanahi for useful discussions and M. Flory for reading the draft before submission and many useful comments.  This work has been supported by National Elites Foundation (INEF) of Iran.

\appendix 

\section{Higher orders of complexity}\label{appendix}

In this appendix we expand the complexity in the second order. So first reviewing the steps in \cite{Carmi:2016wjl}, we write the induced metric on the embedded hypersurface, $H$ as
\begin{gather}
h_{AB}= e_A^\mu \ e_B^\nu \ G_{\mu\nu}= \partial_A X^\mu \ \partial_B X^\nu \ G_{\mu\nu}.
\end{gather}

There is a normal vector on $H$, $n^\mu$, which we normalize as $G_{\mu\nu} n^\mu n^\nu \equiv \epsilon = \pm 1$, depending whether $H$ is timelike or spacelike.  The profile of $H$ is specified through the ``Gauss-Weingarten" equation
\begin{gather}
\partial_A \partial_B X^\mu + {\Gamma^\mu}_{\alpha \beta} \ \partial_A X^\alpha \ \partial_B X^\beta -{\gamma^C}_{AB} \ \partial_C X^\mu= -\epsilon \ K_{AB} \ n^\mu
\end{gather}
where ${\Gamma^\mu}_{\alpha \beta}$ and ${\gamma^C}_{AB}$ are constructed from $G_{\mu\nu}$ and $h_{AB}$, respectively and $K_{AB}$ is the extrinsic curvature on $H$.  If $H$ is extremal, then $K=\text{Tr} \ K_{AB}=0$ and thus the Gauss-Weingarten equation reduces to 
\begin{gather}\label{eq:GW eq}
h^{AB}\  \partial_A \partial_B X^\mu+h^{AB} \ {\Gamma^\mu}_{\alpha \beta} \ \partial_A X^\alpha \ \partial_B X^\beta-h^{AB} \ {\gamma^C}_{AB} \ \partial_C X^\mu=0. 
\end{gather}

Working in the Fefferman-Graham coordinates, the bulk metric reads 
\begin{gather}
ds^2=\frac{L^2}{4\rho^2} d\rho^2+\frac{1}{\rho} g_{ij}(X,\rho) dX^i dX^j;\nonumber\\
g_{ij} (X,\rho)= g_{ij}^{(0)} (X)+\rho \ g_{ij}^{(1)} (X)+\rho^2 g_{ij}^{(2)} (X)+  \dots + \rho^{\frac{d}{2}} g_{ij} ^{(\frac{d}{2})} (X) +\rho^{\frac{d}{2}}\  \log \rho \ f_{ij}(X)+\dots
\end{gather}
and logarithm appears in only even dimensions. 

The induced metric on $H$ finds the following expansion in $\rho$
\[
    h_{AB}=\left\{
                \begin{array}{ll}
                  h_{\rho \rho }= \frac{L^2}{4\rho^2} \left (1+\rho \ h_{\rho \rho}^{(1)} + \dots \right),  \\
                  h_{ab} =\frac{1}{\rho}  \left( h_{ab}^{(0)}+\rho \ h_{ab}^{(1)}+\dots \right).  \\
                \end{array}
              \right.
  \]
Note that
\begin{align}
h^{\rho \rho} =& \frac{4 \rho^2}{ L^2} \frac{1}{ 1+\rho \ h_{\rho \rho} ^{(1)}}=\frac{4 \rho^2}{L^2} \left(1-\rho \ h_{\rho \rho}^{(1)} +\dots \right), \nonumber\\
h^{ab}=& \rho \left( h^{(0) ab} -\rho \ h^{(1) ab} +\dots \right),
\end{align}
where 
\begin{gather}
h^{(1) ab}=h ^{(0) ac} \ h^{(0) bd} \ h_{cd} ^{(1)}. 
\end{gather}
The desired components of Gauss-Weingarten equation is then
\begin{gather}
h^{ab} \ \partial_a   \partial_b   X^i +h^{ab} \  \Gamma_{jk}^i  \ \partial_a X^j \  \partial_b X^k  - h^{ab} \  \gamma_{ab}^C \  \partial_C X^i=0.
\end{gather}
The last term can be decomposed to
\begin{gather}
h^{ab} \  \gamma^C_{ab} \  \partial_C X^i= h^{ab} \  \gamma _{ab}^\rho \ \partial_\rho X^i +h^{ab} \  \gamma^c_{ab} \  \partial_c X^i,
\end{gather}
but
\begin{gather}
\gamma_{ab}^\rho =-\frac{1}{2} h^{\rho \rho} h_{ab, \rho}= -\frac{1}{2} \times \frac{4\rho^2}{ L^2} \times \left(-\frac{1}{\rho^2} h_{ab}^{(0)} +\dots \right)=\frac{2}{L^2} h_{ab}^{(0)},
\end{gather}
and
\begin{gather}
h^{ab} \ \gamma^\rho_{ab} \ \partial_\rho X^i= \rho \ h^{(0) ab} \ \frac{2}{L^2}  \ h^{(0)}_{ab}\  \partial_\rho X^i=\rho \ \frac{2(d-1)}{L^2} \ \partial_\rho X^i.
\end{gather}

The next step is to expand $X^i$ as follows
\begin{gather}
X^i= X^{(0) i} +\rho \ X^{(1)i}+\dots, 
\end{gather}
Then at the first order of $\rho$ we get
\begin{gather}
\partial^a \ \partial_a X^{(0)i}- h^{(0)ab} \gamma^c_{ab} \partial_c X^{(0) i}+\Gamma_{jk}^i \partial_a X^{(0)j} \partial^a X^{(0)k} =\frac{2(d-1)}{L^2} X^{(1)}.
\end{gather}
So
\begin{gather}
\nabla ^a \partial_a X^{(0)i}+ \Gamma_{jk}^i \ \partial_a X^{(0) j}\ \partial^a X^{(0)k} =\frac{2(d-1)}{L^2} X^{(1)i}.
\end{gather}

The left hand side is by definition the trace of the extrinsic on $\Sigma$, thus
\begin{gather}
X^{(1)i} = - \epsilon \  \frac{L^2}{2 (d-1)} \ K \ n^i,
\end{gather}
where $n^i$ is the unit normal on $\Sigma$ and $K$ is the trace of the extrinsic curvature contracted from $n^i$. 

Now one can compute the expanded induced metric 
\begin{gather}
G_{\mu\nu} \ \partial_\rho X^\mu \ \partial_\rho X^\nu \Longrightarrow \frac{1}{\rho} \  g_{ij} \  \partial_\rho X^i \partial_\rho X^j=\frac{1}{\rho} \ g_{ij}^{(0)} \ X^{(1)i} \ X^{(1)j}\nonumber\\
= \frac{1}{\rho} \  \frac{L^4}{4(d-1)^2} K^2 g_{ij}^{(0)} \ n^i\ n^j =\epsilon \  \frac{1}{\rho} \ \frac{L^4}{4(d-1)^2} K^2,
\end{gather}
and then finds
\begin{gather}
h_{\rho \rho}^{(1)} = \epsilon \ \frac{L^2}{(d-1)^2} \ K^2. 
\end{gather}
Then we can write
\begin{gather}
G_{\mu\nu} \partial_a X^\mu \partial_b X^\nu \xrightarrow {O(\rho)} g_{ij}^{(0)} \left(\partial_a X^{(1)i} \ \partial_b X^{(0) j} + \partial_a X^{(0)i} \partial_b X^{(1)j} \right)+g_{ij}^{(1)} \partial_a X^{(0)i} \partial_b X^{(0)j}, 
\end{gather}
and we have also
\begin{gather} \label{eq:one}
g_{ij}^{(0)} \partial_b X^{(0) j}  \partial_a X^{(1) i} = g_{ij}^{(0)} e_b^j \partial_a \left(-\epsilon \frac{L^2}{2(d-1)} K n^i\right)=-\epsilon \ \frac{L^2}{2(d-1)} g_{ij}^{(0)} \ e_b^j  \left( (\partial_a K) n^i +K \partial_a n^i \right).
\end{gather}
Note that $e.n= g_{ij}^{(0)} e_b^j n^i=0$, thus the first term in \eqref{eq:one} vanishes identically.  For the second term we have
\begin{gather}
g_{ij}^{(0)}   e_b^j \ \partial_a  n^i = e^i_b\  \partial_a  n_i = e_b^i \ e_b ^j  \ n_{i,j} .
\end{gather}
Now since $n.e=0$ we can add and subtract $ \Sigma_{jk}^i n_i$, and therefore we get 
\begin{gather}
g_{ij}^{(0)} \ e_b^j \ \partial_a n^i = e_b^i \ e_b^j \ n_{i;j} =K_{ab},
\end{gather}
therefore
\begin{gather}
g_{ij}^{(0)} \ \partial_b X^{(0)j}  \partial_a X^{(1) i} =- \epsilon \ \frac{L^2}{2(d-1)} \ K  K_{ab}.
\end{gather}

We also know
\begin{gather}
g_{ij}^{(1)} (x)= -\frac{L^2}{d-2} \left( R_{ij}(g^{(0)}) -\frac{ g_{ij}^{(0)} }{2(d-1)}  R(g^{(0)}) \right ),
\end{gather}
so we get
\begin{gather}
g_{ij}^{(1)} (x)  \ \partial_a X^{(0)i} \ \partial_b X^{(0) j} =-\frac{L^2}{d-2} \left( R_{ij} e_a^i \ e_b ^j -\frac{1}{2(d-1)} R(g^{(0)})  g_{ij}^{(0)} \ e_a^i \ e_b^j \right) \nonumber\\
=-\frac{L^2}{d-2} \left( R_{\hat{a} \hat{b}} -\frac{1}{2(d-1)} R \  h_{ab} ^{(0)} \right).
\end{gather}
Therefore we finally get
\begin{gather}
h_{ab}^{(1)} =-\frac{L^2}{ d-1} \left ( \frac{d-1}{d-2} \ R_{\hat{a} \hat{b}} -\frac{1}{2(d-2)} R \  h_{ab}^{(0)} +\epsilon  K  K_{ab} \right).
\end{gather}
Now we want to calculate the volume
\begin{gather}
V=\int d^d x \sqrt{h} \ \Big |_{\text{extremal}},
\end{gather}
where
\begin{gather}
\sqrt{h}= \sqrt{h_{\rho \rho}} \sqrt{h_{ab}}\ ,  \ \ \ \ \ \ \ \ \ \ \sqrt{h_{\rho \rho}}=\frac{L}{2\rho} \left(1+\frac{1}{2} \  \rho \ h_{\rho \rho} ^{(1)}\right). 
\end{gather}
We can also use the lemma that if $g_{ij} = \bar{g_{ij}}+\epsilon h_{ij}$, then
\begin{gather}
\sqrt{g} =\sqrt{\bar{g}} \left(1+\frac{1}{2} \epsilon h_i^i -\frac{1}{4} \epsilon^2 h_{ij} h^{ij}+\frac{1}{8} \epsilon^2 (h_i^i)^2 \right)+\dots 
\end{gather}
where $h_i^i=\bar{g}^{ij} h_{ij}$ and $h_{ij} h^{ij} = \bar {g}^{ik} \bar{g}^{jl} h_{ij} h_{kl}$.
 Additionally, 
\begin{gather}
\sqrt{ | h_{ab} |}=\frac{1}{\rho ^{\frac{d-1}{2}}} \sqrt{h^{(0)}} \left (1+\frac{1}{2} \  \rho  \ h^{(0) ab} h_{ab}^{(1)}+\dots \right),
\end{gather}
so
\begin{gather}
\sqrt{h}=\frac{L}{2 \rho^{\frac{d+1}{2}}} \sqrt{h^{(0)}}  \left(1+\frac{1}{2} \rho (h_{\rho \rho}^{(1)} + h^{(0) ab} h_{ab} ^{(1)} ) \right)+\dots ,
\end{gather}
then
\begin{gather}
h_{ab}^{(0)} h^{(1) ab}=-\frac{L^2}{ d-1} \left(\frac{d-1}{d-2} R_{\hat{a}}^{\hat{a}}-\frac{1}{2(d-2)} R (d-1) +\epsilon K^2 \right).
\end{gather}
Therefore,
\begin{gather}
V=\int d^{d-1} x \sqrt{h^{(0)}}  \ \int_{\frac{\delta^2}{L^2}} d\rho \frac{L}{2 \rho^{\frac{d+1}{2}} } \left [ 1+\frac{1}{2}\rho  \left (\epsilon \frac{L^2}{(d-1)^2} K^2-\frac{L^2}{d-1} \left(\frac{d-1}{d-2} R_{\hat{a}}^{\hat{a}} -\frac{d-1}{2(d-2)} R+\epsilon K^2 \right) \right) \right]\nonumber\\
=\int d^{d-1} x \sqrt{h^{(0)}} \int_{\frac{\delta^2}{L^2}} d\rho \frac{L}{2\rho^{\frac{d+1}{2}}}+
\int d^{d-1} x \sqrt{h^{(0)}} \int_{\frac{\delta^2}{L^2}} d\rho \frac{L^3}{4 (d-2) \rho^{\frac{d-1}{2}}} \left(-R_{\hat{a}}^{\hat{a}}+\frac{1}{2} R -\frac{(d-2)^2}{(d-1)^2} \epsilon K^2 \right)\nonumber\\
=\frac{L^d}{d-1} \int d^{d-1} x \sqrt{h^{(0)}} \left(\frac{1}{\delta^{d-1}}-\frac{d-1}{2 (d-2)(d-3) \delta^{d-3}} \left(R_{\hat{a}}^ {\hat{a}}-\frac{1}{2} R +\epsilon \ \frac{(d-2)^2}{(d-1)^2} K^2 \right)\right).
\end{gather}
For $d=3$, we get a logarithm as
\begin{gather}
V_{d=3}^{\text{log}} =\frac{L^3}{8} \log \left(\frac{\delta}{L}\right) \int d^2 x \sqrt{h^{(0)}} \left( 4 R_{\hat{a}}^{\hat{a}}-2R+\epsilon K^2 \right).
\end{gather}
which is the result in  \cite{Carmi:2016wjl}.

Now it would be interesting to perform the similar procedure in the second order of the metric and then may even higher orders to see if one can get more universal results in the different terms of the expansions of complexity.

So first, in the second order one would have
\begin{flalign}
\gamma_{ab}^\rho &= \frac{2}{L^2} h_{ab}^{(0)}-\frac{2 \rho^2}{L^2} h_{\rho \rho}^{(1)} h_{ab}^{(0)}-\frac{2 \rho^2}{L^2} h_{ab}^{(0)} h_{\rho \rho}^{(2)}-\frac{2\rho^2}{L^2} h_{ab}^{(2)}.
\end{flalign}
Then from \eqref{eq:GW eq}, for the second order of $\rho$ one would get
\begin{gather}
X^{(2)i}=\frac{\epsilon L^2}{4} K n^i h^{(1) ab} -\Big (\frac{d-1}{2}h_{\rho \rho}^{(1)}  -\frac{1}{2} h_{ab}^{(0)}  h^{(1) ab} \Big) X^{(1) i}  + \nonumber\\ \frac{L^2}{4} \bigg[ \nabla^a \partial_a X^{(1) i}+\Gamma_{jk}^i h^{(0) ab} \big( \partial_a X^{(0) j} \partial_b X^{(1) k} +\partial_a X^{(1)j} \partial_b X^{(0) k} \big) \bigg].
\end{gather}
Then $h_{\rho \rho}^{(2)}$ would be
\begin{gather}
h_{\rho \rho}^{(2)} =\frac{L^4 K^2}{\rho (d-1)^2 (d-2)} \left ( R_{ij}^{(0)} n^i n^j -\frac{R(g^{(0)})}{2(d-1)} \right)-\left( \frac{8 \epsilon }{d-1}\right) K n_j X^{(2) j}.
\end{gather}

For finding the second order of $h_{ab}^{(2)}$ we should find the following terms
\begin{gather}
G_{\mu \nu} \partial_a X^\mu \partial_b X^\nu \Longrightarrow \underbrace{g_{ij}^{(0)} \partial_a X^{(1)i} \partial_b X^{(1) j} }_{(1)}+ \underbrace{ g_{ij}^{(1)} \Big( \partial_a X^{(1)i} \partial_b X^{(0) j} +\partial_a X^{(0) i} \partial_b X^{(1) j} \Big)}_{(2)}+\underbrace{g_{ij}^{(2)} \partial_a X^{(0) i} \partial_b X^{(0) j}}_{(3)}. 
\end{gather}
The first term would be
\begin{gather}
(1) \to \frac{g_{ij}^{(0)} L^4  }{4(d-1)^2} \Big( (\partial_a K)(\partial_b K)n^i n^j +K^2 (\partial_a n^i)(\partial_b n^j) +K\left(\partial_a K n^i \partial_b n^j +\partial_b K n^j \partial_a n^i\right) \Big).
\end{gather}
The second term is
\begin{gather}
(2) \to \frac{\epsilon L^4}{2(d-1) (d-2)}  \Big[ R_{ij}(g^{(0)})  \big( \partial_a K n^i e_b^j+\partial_b K n^j e_a^i+K \partial_a n^i e_b^j+K \partial_b n^j e_a^i \big)\nonumber\\
-\frac{R(g^{(0)}) K}{2(d-1)} \left( \partial_a n_j e_b^j+\partial_b n_j e_a^i\right) \Big],
\end{gather}
and the third term is 
\begin{gather}
(3) \to g_{ij}^{(2)} e_a^i e_b^j.
\end{gather}
where from the appendix of \cite{deHaro:2000vlm}, $g_{ij}^{(2)}$ is
\begin{gather}
g_{ij}^{(2)}=\frac{1}{d-4} \Big( -\frac{1}{8(d-1)} D_i D_j R+\frac{1}{4(d-2)} D_k D^k R_{ij}
-\frac{1}{8(d-1)(d-2)} D_k D^k R g_{ij}^{(0)}-\frac{1}{2(d-2)} R^{kl} R_{ikjl}\nonumber\\
+\frac{d-4}{2(d-2)^2} R_i^k R_{kj} +\frac{1}{(d-1)(d-2)^2} R R_{ij}
+\frac{1}{4(d-2)^2} R^{kl} R_{kl} g_{ij}^{(0)} -\frac{3d}{16 (d-1)^2 (d-2)^2} R^2 g_{ij}^{(0)} \Big).
\end{gather}

 \medskip

\bibliography{complex}

\providecommand{\href}[2]{#2}\begingroup\raggedright\begin{thebibliography}{10}

\bibitem{Watrous:2008}
J.~Watrous, {\it Quantum computational complexity},  {\em ncyclopedia of
  Complexity and Systems Science,ed., R. A. Meyers} (2009)
  [\href{http://xxx.lanl.gov/abs/quant-ph"0804.3401}{{\tt
  quant-ph"0804.3401}}].

\bibitem{Chapman:2017rqy}
S.~Chapman, M.~P. Heller, H.~Marrochio, and F.~Pastawski, {\it {Towards
  Complexity for Quantum Field Theory States}},
  \href{http://xxx.lanl.gov/abs/1707.08582}{{\tt arXiv:1707.08582}}.

\bibitem{Jefferson:2017sdb}
R.~A. Jefferson and R.~C. Myers, {\it {Circuit complexity in quantum field
  theory}},  \href{http://xxx.lanl.gov/abs/1707.08570}{{\tt arXiv:1707.08570}}.

\bibitem{Susskind:2014rva}
L.~Susskind, {\it {Computational Complexity and Black Hole Horizons}},  {\em
  Fortsch. Phys.} {\bf 64} (2016) 44--48,
  [\href{http://xxx.lanl.gov/abs/1403.5695}{{\tt arXiv:1403.5695}}]. [Fortsch.
  Phys.64,24(2016)].

\bibitem{Brown:2015bva}
A.~R. Brown, D.~A. Roberts, L.~Susskind, B.~Swingle, and Y.~Zhao, {\it
  {Holographic Complexity Equals Bulk Action?}},  {\em Phys. Rev. Lett.} {\bf
  116} (2016), no.~19 191301, [\href{http://xxx.lanl.gov/abs/1509.07876}{{\tt
  arXiv:1509.07876}}].

\bibitem{Lloyd:2000}
S.~Lloyd, {\it Ultimate physical limits to computation},  {\em Nature 406
  (2000) 1047} (2000) [\href{http://xxx.lanl.gov/abs/quant-ph"/9908043}{{\tt
  quant-ph"/9908043}}].

\bibitem{Brown:2015lvg}
A.~R. Brown, D.~A. Roberts, L.~Susskind, B.~Swingle, and Y.~Zhao, {\it
  {Complexity, action, and black holes}},  {\em Phys. Rev.} {\bf D93} (2016),
  no.~8 086006, [\href{http://xxx.lanl.gov/abs/1512.04993}{{\tt
  arXiv:1512.04993}}].

\bibitem{Alishahiha:2017hwg}
M.~Alishahiha, A.~Faraji~Astaneh, A.~Naseh, and M.~H. Vahidinia, {\it {On
  complexity for F(R) and critical gravity}},  {\em JHEP} {\bf 05} (2017) 009,
  [\href{http://xxx.lanl.gov/abs/1702.06796}{{\tt arXiv:1702.06796}}].

\bibitem{Hughes:2015ora}
T.~L. Hughes, R.~G. Leigh, O.~Parrikar, and S.~T. Ramamurthy, {\it
  {Entanglement entropy and anomaly inflow}},  {\em Phys. Rev.} {\bf D93}
  (2016), no.~6 065059, [\href{http://xxx.lanl.gov/abs/1509.04969}{{\tt
  arXiv:1509.04969}}].

\bibitem{Iqbal:2015vka}
N.~Iqbal and A.~C. Wall, {\it {Anomalies of the Entanglement Entropy in Chiral
  Theories}},  {\em JHEP} {\bf 10} (2016) 111,
  [\href{http://xxx.lanl.gov/abs/1509.04325}{{\tt arXiv:1509.04325}}].

\bibitem{Castro:2015csg}
A.~Castro, D.~M. Hofman, and N.~Iqbal, {\it {Entanglement Entropy in Warped
  Conformal Field Theories}},  {\em JHEP} {\bf 02} (2016) 033,
  [\href{http://xxx.lanl.gov/abs/1511.00707}{{\tt arXiv:1511.00707}}].

\bibitem{Miskovic:2009kr}
O.~Miskovic and R.~Olea, {\it {Background-independent charges in Topologically
  Massive Gravity}},  {\em JHEP} {\bf 12} (2009) 046,
  [\href{http://xxx.lanl.gov/abs/0909.2275}{{\tt arXiv:0909.2275}}].

\bibitem{Bouchareb:2007yx}
A.~Bouchareb and G.~Clement, {\it {Black hole mass and angular momentum in
  topologically massive gravity}},  {\em Class. Quant. Grav.} {\bf 24} (2007)
  5581--5594, [\href{http://xxx.lanl.gov/abs/0706.0263}{{\tt
  arXiv:0706.0263}}].

\bibitem{Alishahiha:2015rta}
M.~Alishahiha, {\it {Holographic Complexity}},  {\em Phys. Rev.} {\bf D92}
  (2015), no.~12 126009, [\href{http://xxx.lanl.gov/abs/1509.06614}{{\tt
  arXiv:1509.06614}}].

\bibitem{Grumiller:2008qz}
D.~Grumiller and N.~Johansson, {\it {Instability in cosmological topologically
  massive gravity at the chiral point}},  {\em JHEP} {\bf 07} (2008) 134,
  [\href{http://xxx.lanl.gov/abs/0805.2610}{{\tt arXiv:0805.2610}}].

\bibitem{Zhang:2013mva}
B.~Zhang, {\it {Statistical entropy of a BTZ black hole in topologically
  massive gravity}},  {\em Phys. Rev.} {\bf D88} (2013) 124017,
  [\href{http://xxx.lanl.gov/abs/1307.7639}{{\tt arXiv:1307.7639}}].

\bibitem{Myung:2008ey}
Y.~S. Myung, H.~W. Lee, and Y.-W. Kim, {\it {Entropy of black holes in
  topologically massive gravity}},
  \href{http://xxx.lanl.gov/abs/0806.3794}{{\tt arXiv:0806.3794}}.

\bibitem{Flory:2017ftd}
M.~Flory, {\it {A complexity/fidelity susceptibility $g$-theorem for
  AdS$_{3}$/BCFT$_{2}$}},  {\em JHEP} {\bf 06} (2017) 131,
  [\href{http://xxx.lanl.gov/abs/1702.06386}{{\tt arXiv:1702.06386}}].

\bibitem{Roy:2017uar}
P.~Roy and T.~Sarkar, {\it {On subregion holographic complexity and
  renormalization group flows}},
  \href{http://xxx.lanl.gov/abs/1708.05313}{{\tt arXiv:1708.05313}}.

\bibitem{Ferreira:2013zta}
H.~R.~C. Ferreira, {\it {Stability of warped AdS3 black holes in Topologically
  Massive Gravity under scalar perturbations}},  {\em Phys. Rev.} {\bf D87}
  (2013), no.~12 124013, [\href{http://xxx.lanl.gov/abs/1304.6131}{{\tt
  arXiv:1304.6131}}].

\bibitem{Chen:2009hg}
B.~Chen and Z.-b. Xu, {\it {Quasi-normal modes of warped black holes and warped
  AdS/CFT correspondence}},  {\em JHEP} {\bf 11} (2009) 091,
  [\href{http://xxx.lanl.gov/abs/0908.0057}{{\tt arXiv:0908.0057}}].

\bibitem{Freedman:2016zud}
M.~Freedman and M.~Headrick, {\it {Bit threads and holographic entanglement}},
  {\em Commun. Math. Phys.} {\bf 352} (2017), no.~1 407--438,
  [\href{http://xxx.lanl.gov/abs/1604.00354}{{\tt arXiv:1604.00354}}].

\bibitem{Caputa:2017yrh}
P.~Caputa, N.~Kundu, M.~Miyaji, T.~Takayanagi, and K.~Watanabe, {\it {Liouville
  Action as Path-Integral Complexity: From Continuous Tensor Networks to
  AdS/CFT}},  \href{http://xxx.lanl.gov/abs/1706.07056}{{\tt
  arXiv:1706.07056}}.

\bibitem{Anninos:2008fx}
D.~Anninos, W.~Li, M.~Padi, W.~Song, and A.~Strominger, {\it {Warped AdS(3)
  Black Holes}},  {\em JHEP} {\bf 03} (2009) 130,
  [\href{http://xxx.lanl.gov/abs/0807.3040}{{\tt arXiv:0807.3040}}].

\bibitem{Dereli:2000fm}
T.~Dereli and O.~Sarioglu, {\it {Topologically massive gravity and black holes
  in three-dimensions}},  \href{http://xxx.lanl.gov/abs/gr-qc/0009082}{{\tt
  gr-qc/0009082}}.

\bibitem{Moussa:2008sj}
K.~A. Moussa, G.~Clement, H.~Guennoune, and C.~Leygnac, {\it {Three-dimensional
  Chern-Simons black holes}},  {\em Phys. Rev.} {\bf D78} (2008) 064065,
  [\href{http://xxx.lanl.gov/abs/0807.4241}{{\tt arXiv:0807.4241}}].

\bibitem{Ghodrati:2017rta}
M.~Ghodrati, {\it {to be submitted}}, .

\bibitem{Alishahiha:2016cjk}
M.~Alishahiha, A.~Davody, A.~Naseh, and S.~F. Taghavi, {\it {On Butterfly
  effect in Higher Derivative Gravities}},  {\em JHEP} {\bf 11} (2016) 032,
  [\href{http://xxx.lanl.gov/abs/1610.02890}{{\tt arXiv:1610.02890}}].

\bibitem{Hohm:2010jc}
O.~Hohm and E.~Tonni, {\it {A boundary stress tensor for higher-derivative
  gravity in AdS and Lifshitz backgrounds}},  {\em JHEP} {\bf 04} (2010) 093,
  [\href{http://xxx.lanl.gov/abs/1001.3598}{{\tt arXiv:1001.3598}}].

\bibitem{Ghodrati:2016vvf}
M.~Ghodrati, K.~Hajian, and M.~R. Setare, {\it {Revisiting Conserved Charges in
  Higher Curvature Gravitational Theories}},  {\em Eur. Phys. J.} {\bf C76}
  (2016), no.~12 701, [\href{http://xxx.lanl.gov/abs/1606.04353}{{\tt
  arXiv:1606.04353}}].

\bibitem{Oliva:2009ip}
J.~Oliva, D.~Tempo, and R.~Troncoso, {\it {Three-dimensional black holes,
  gravitational solitons, kinks and wormholes for BHT massive gravity}},  {\em
  JHEP} {\bf 07} (2009) 011, [\href{http://xxx.lanl.gov/abs/0905.1545}{{\tt
  arXiv:0905.1545}}].

\bibitem{Nam:2010dd}
S.~Nam, J.-D. Park, and S.-H. Yi, {\it {AdS Black Hole Solutions in the
  Extended New Massive Gravity}},  {\em JHEP} {\bf 07} (2010) 058,
  [\href{http://xxx.lanl.gov/abs/1005.1619}{{\tt arXiv:1005.1619}}].

\bibitem{Giribet:2009qz}
G.~Giribet, J.~Oliva, D.~Tempo, and R.~Troncoso, {\it {Microscopic entropy of
  the three-dimensional rotating black hole of BHT massive gravity}},  {\em
  Phys. Rev.} {\bf D80} (2009) 124046,
  [\href{http://xxx.lanl.gov/abs/0909.2564}{{\tt arXiv:0909.2564}}].

\bibitem{Capela:2012uk}
F.~Capela and G.~Nardini, {\it {Hairy Black Holes in Massive Gravity:
  Thermodynamics and Phase Structure}},  {\em Phys. Rev.} {\bf D86} (2012)
  024030, [\href{http://xxx.lanl.gov/abs/1203.4222}{{\tt arXiv:1203.4222}}].

\bibitem{Ghodrati:2016ggy}
M.~Ghodrati and A.~Naseh, {\it {Phase transitions in Bergshoeff-Hohm-Townsend
  massive gravity}},  {\em Class. Quant. Grav.} {\bf 34} (2017), no.~7 075009,
  [\href{http://xxx.lanl.gov/abs/1601.04403}{{\tt arXiv:1601.04403}}].

\bibitem{Ghodrati:2016tdy}
M.~Ghodrati, {\em {Beyond AdS Space-times, New Holographic Correspondences and
  Applications}}.
\newblock PhD thesis, Michigan U., MCTP, 2016.
\newblock \href{http://xxx.lanl.gov/abs/1609.04168}{{\tt arXiv:1609.04168}}.

\bibitem{Clement:2009ka}
G.~Clement, {\it {Black holes with a null Killing vector in new massive gravity
  in three dimensions}},  {\em Class. Quant. Grav.} {\bf 26} (2009) 165002,
  [\href{http://xxx.lanl.gov/abs/0905.0553}{{\tt arXiv:0905.0553}}].

\bibitem{Ghodsi:2010gk}
A.~Ghodsi and M.~Moghadassi, {\it {Charged Black Holes in New Massive
  Gravity}},  {\em Phys. Lett.} {\bf B695} (2011) 359--364,
  [\href{http://xxx.lanl.gov/abs/1007.4323}{{\tt arXiv:1007.4323}}].

\bibitem{Ghodrati:2014spa}
M.~Ghodrati, {\it {Hyperscaling Violating Solution in Coupled Dilaton-Squared
  Curvature Gravity}},  {\em Phys. Rev.} {\bf D90} (2014), no.~4 044055,
  [\href{http://xxx.lanl.gov/abs/1404.5399}{{\tt arXiv:1404.5399}}].

\bibitem{Sachs:2011xa}
I.~Sachs, {\it {Formation of black holes in topologically massive gravity}},
  {\em Phys. Rev.} {\bf D87} (2013), no.~2 024019,
  [\href{http://xxx.lanl.gov/abs/1108.3579}{{\tt arXiv:1108.3579}}].

\bibitem{Flory:2013ssa}
M.~Flory and I.~Sachs, {\it {Dynamical Black Holes in 2+1 Dimensions}},  {\em
  Phys. Rev.} {\bf D88} (2013) 044034.

\bibitem{Carmi:2016wjl}
D.~Carmi, R.~C. Myers, and P.~Rath, {\it {Comments on Holographic Complexity}},
   {\em JHEP} {\bf 03} (2017) 118,
  [\href{http://xxx.lanl.gov/abs/1612.00433}{{\tt arXiv:1612.00433}}].

\bibitem{Alishahiha:2017cuk}
M.~Alishahiha and A.~Faraji~Astaneh, {\it {Holographic Fidelity
  Susceptibility}},  \href{http://xxx.lanl.gov/abs/1705.01834}{{\tt
  arXiv:1705.01834}}.

\bibitem{Hashimoto:2017fga}
K.~Hashimoto, N.~Iizuka, and S.~Sugishita, {\it {Time Evolution of Complexity
  in Abelian Gauge Theories - And Playing Quantum Othello Game -}},
  \href{http://xxx.lanl.gov/abs/1707.03840}{{\tt arXiv:1707.03840}}.

\bibitem{Nielsen:2005}
M.~A. Nielsen, {\it {A geometric approach to quantum circuit lower bounds}},
  \href{http://xxx.lanl.gov/abs/0502070}{{\tt 0502070}}.

\bibitem{Brown:2017jil}
A.~R. Brown and L.~Susskind, {\it {The Second Law of Quantum Complexity}},
  \href{http://xxx.lanl.gov/abs/1701.01107}{{\tt arXiv:1701.01107}}.

\bibitem{Brown:2016wib}
A.~R. Brown, L.~Susskind, and Y.~Zhao, {\it {Quantum Complexity and Negative
  Curvature}},  {\em Phys. Rev.} {\bf D95} (2017), no.~4 045010,
  [\href{http://xxx.lanl.gov/abs/1608.02612}{{\tt arXiv:1608.02612}}].

\bibitem{Czech:2017ryf}
B.~Czech, {\it {Einstein's Equations from Varying Complexity}},
  \href{http://xxx.lanl.gov/abs/1706.00965}{{\tt arXiv:1706.00965}}.

\bibitem{Compere:2013aya}
G.~Compère, W.~Song, and A.~Strominger, {\it {Chiral Liouville Gravity}},  {\em
  JHEP} {\bf 05} (2013) 154, [\href{http://xxx.lanl.gov/abs/1303.2660}{{\tt
  arXiv:1303.2660}}].

\bibitem{Bao:2016rbj}
N.~Bao and A.~Chatwin-Davies, {\it {The Complexity of Identifying
  Ryu-Takayanagi Surfaces in AdS3/CFT2}},  {\em JHEP} {\bf 11} (2016) 034,
  [\href{http://xxx.lanl.gov/abs/1609.01727}{{\tt arXiv:1609.01727}}].

\bibitem{Bhattacharya:2014vja}
J.~Bhattacharya, V.~E. Hubeny, M.~Rangamani, and T.~Takayanagi, {\it
  {Entanglement density and gravitational thermodynamics}},  {\em Phys. Rev.}
  {\bf D91} (2015), no.~10 106009,
  [\href{http://xxx.lanl.gov/abs/1412.5472}{{\tt arXiv:1412.5472}}].

\bibitem{Ghodrati:2015rta}
M.~Ghodrati, {\it {Schwinger Effect and Entanglement Entropy in Confining
  Geometries}},  {\em Phys. Rev.} {\bf D92} (2015), no.~6 065015,
  [\href{http://xxx.lanl.gov/abs/1506.08557}{{\tt arXiv:1506.08557}}].

\bibitem{deHaro:2000vlm}
S.~de~Haro, S.~N. Solodukhin, and K.~Skenderis, {\it {Holographic
  reconstruction of space-time and renormalization in the AdS / CFT
  correspondence}},  {\em Commun. Math. Phys.} {\bf 217} (2001) 595--622,
  [\href{http://xxx.lanl.gov/abs/hep-th/0002230}{{\tt hep-th/0002230}}].

\end{thebibliography}\endgroup
\bibliographystyle{JHEP}
\end{document}